# Optimal control in opinion dynamics models: towards a unified framework

## Ivan V. Kozitsin

Laboratory of Active Systems, V. A. Trapeznikov Institute of Control Sciences of Russian Academy of Sciences, 65 Profsoyuznaya street, Moscow, 117997, Russian Federation

Department of Higher Mathematics, Moscow Institute of Physics and Technology, 9 Institutskiy per., Dolgoprudny, Moscow Region, 141701, Russian Federation

Email: kozitsin.ivan@mail.ru

Understanding how individuals change their opinions is essential to mitigate the harmful effect of fake news and stop opinion polarization. Building upon an extremely flexible opinion dynamics model that can capture the key microscopic mechanisms of social influence, the current paper elaborates a framework to find optimal control, which, thus, can be applied to a broad set of opinion dynamics settings. Using a combination of theoretical and computational approaches, I characterize the properties of optimal control and demonstrates how the elaborated framework can be applied to specific examples, which cover the classical situations that garner remarkable attention in the literature on opinion dynamics models: assimilative influence, bounded confidence, and dissimilative influence.

## 1. Introduction

The growing popularity of online communication services has made scholars pay more attention to how information diffuses through the Internet, how users change their behaviour, and why fake news prevails in the online environment. A crucial backbone behind answering all these questions is so-called opinion formation models (aka social influence models) that were within the scope of scientific interest since approximately the mid-twentieth century (French Jr, 1956; Rashevsky, 1939). Opinion formation models concern how individuals change their opinions while communicating with each other and receiving different sorts of information (including those coming from the Internet). To date, a huge amount of opinion formation models has been introduced. Perfect reviews of them are presented in (Castellano *et al*., 2009; Flache *et al*., 2017; Mastroeni *et al*., 2019; Proskurnikov & Tempo, 2017, 2018). In a nutshell, the current situation in this field is that the theoretical proliferation of social influence models prevails over their practical applications to real social processes (Chattoe-Brown, 2020; Vazquez, 2022). A comprehensive study that shed light on the problems regarding validations of these models can be found in (Mäs, 2019). Interestingly, as was pointed in (Chattoe-Brown, 2022), those papers that introduce opinion formation models and skip the validation step tend to get more citations, compared to those that try also to validate the proposed model (however, this analysis was restricted by the papers published in only one journal *JASSS*[1]). Of course, without validation, one cannot say what social influence mechanism stands behind opinion formation processes (Chan *et al.*, 2022; Mei *et al.*, 2022).

As was pointed out in (Friedkin, 2015), "The coordination and control of social systems is the foundational problem of sociology." The recent COVID-19 pandemic has demonstrated how it is important to consolidate society by following government recommendations of wearing facemasks, being vaccinated, and keeping social distance. In this regard, knowing the micro-foundations of opinion formation processes is necessary to develop scientifically-grounded control strategies that can help convince citizens of this sort of behaviour. Further examples that demonstrate the importance of control over social systems refer to promoting healthy behaviour (Aral & Nicolaides, 2017), mitigating the spread of fake news (Khan *et al*., 2021), and stopping opinion polarization (Balietti *et al.*, 2021; Li & Zhao, 2021).

---

[1] https://www.jasss.org/JASSS.html

At the time, all attempts to develop control strategies applied to social systems of interacting opinions have a rather non-systematic character. Typically, scholars chose some special opinion formation model (which, according to the authors' intuition, describes better real social processes than other ones) that serves as a basis for the elaboration of optimal control (whatever the objective is) (Brede *et al.*, 2018; Dietrich *et al.*, 2018; Kurz, 2015; Li & Zhao, 2021; Moreno *et al.*, 2021; Varma *et al.*, 2021). However, because one does not know exactly the organization of the social influence mechanism that governs opinion dynamics, one cannot guarantee that the control derived within the framework of some opinion formation model will lead to the fulfilment of the objectives in practice. Further, the concrete approach of obtaining the optimal control derived within some opinion formation model may not be applicable for other models just because these models are built upon completely different mathematical constructions.

Recently, in (Kozitsin, 2022), the author proposed what he called "a general framework" that aims to model opinion formation processes and bring together theoretical and practical sides of the field. That elaborated on a model which can cover a huge scope of microscopic mechanisms of social influence, including conformism, anticonformism, assimilative influence, bounded confidence, and repulsive (aka dissimilar) influence (Abramiuk-Szurlej *et al.*, 2021; Carro *et al.*, 2016; Krueger *et al.*, 2017; Takács *et al.*, 2016). Although this model still cannot approximate some important mechanisms including, for example, threshold effects (Granovetter, 1978), the great flexibility of the model indicates that it could be a nice workhorse based on which one can formulate and solve control problems. Importantly, the resulting solutions will potentially have a wide area of application. The current paper is dedicated to this very purpose.

The contributions of this paper are as follows. First, it presents a generalization of the model (Kozitsin, 2022) which includes stubborn agents who do not change their own opinions but can influence other individuals. Next, for the generalized model, under assumptions of the huge number of agents and the completely mixed population, I develop a mean-field approximation that describes the social system via a set of ordinary differential equations, in which opinions of stubborn agents (presented by macro-variables – as well as opinions of ordinary agents) are exogenous to the model. After that, a control problem is formulated, in which one should adjust what opinions stubborn agents have to disseminate in order to maximize a functional objective. For the control problem, I apply the Pontryagin maximum principle and derives some theoretical results from it. Notably, the organization of the optimal control is characterized. The control problem is solved numerically, by using the forward-backward sweep method. The paper demonstrates how this algorithm works through stylized examples, in which the key microscopic mechanisms of social influence are covered (assimilative influence, bounded confidence, and dissimilative influence) (Flache *et al.*, 2017). Next, the mean-field predictions (and control strategies drawn upon them) are compared with the behaviour of the model in the case of the structured populations (characterized by complex non-complete social graphs). The paper ends up with a conclusion, which, along the way, formulates possible directions for future studies.

## 2. Preliminaries

Within this paper, $[n]$ is used to denote the set of all natural numbers ($\mathbb{N}$) from one to $n \in \mathbb{N}$. The set of real numbers is presented as $\mathbb{R}$. The cardinal number of a set is signified by $\#\{\dots\}$. The notation $\Pr\{\dots\}$ symbolizes the probability of some event. Vectors are denoted with an arrow like this: $\vec{v}$. The inequality $\vec{v} \geq \vec{0}$ indicates that all the components of $\vec{v}$ are nonnegative. In turn, the inequality $\vec{v} \geq \vec{w}$ means that $\vec{v} - \vec{w} \geq \vec{0}$. Matrices are referred to with capital letters, but not all capital letters stand for matrices: they could be also natural numbers or some sets. Throughout this paper, 3-D matrices are broadly used. To depict them, their slices over the first axis are usually employed. For a 3-D matrix $P \in \mathbb{R}^{m \times m \times m}$, its slices over the first axis are denoted as follows: $P_1, \dots, P_m$. The matrices $P_1, \dots, P_m$ are 2-D square matrices that belong to the space $\mathbb{R}^{m \times m}$. In below, summation operators will be broadly used. Unless the other specified, summation indices in that sums will go through the set $[m]$.

### 3. Opinion dynamics model

*3.1. Model setup*

The model concerns the opinion dynamics of $N$ agents connected by a social network $G = (V, E)$, where $V$ corresponds to the nodes of the network and $E$ represents ties between them. The connections are assumed to be undirected and unweighted. Self-loops are not allowed. At each time moment $t$, a randomly chosen agent $i$ is influenced by one of its neighbours – say, agent $j$ – who is also chosen by chance. As a result of communication, agent $i$'s opinion $o_i(t)$ is changed subject to $j$'s position $o_j(t)$. Agents' opinions belong to the discrete opinion space $X = \{x_1, \ldots, x_m\}$, whose entries may be arranged depending on a particular context: $x_1 \prec x_2 \prec \cdots \prec x_m$. The variable $m$ represents the number of possible opinion values. Without loss of generality, it can be assumed that $o_i(t) = x_s, o_j(t) = x_l$. Agent $i$'s new opinion appears according to the distribution $\{p_{s,l,1}, \ldots, p_{s,l,m}\}$ in which the quantity $p_{s,l,k}$ is defined as follows:

$$p_{s,l,k} = \Pr\{o_i(t+1) = x_k \mid o_i(t) = x_s, o_j(t) = x_l\}.$$

To put it simply, $p_{s,l,k}$ is the probability of opinion change $x_s \rightarrow x_k$ given the influence coming from opinion $x_l$. The quantities $p_{s,l,k}$ form the 3D matrix $P = \left[p_{s,l,k}\right]_{s,l,k=1}^{m} \in \mathbb{R}^{m \times m \times m}$. The organization of the matrix $P$ is that its slices over the first axis $P_1, \ldots, P_m \in \mathbb{R}^{m \times m}$ are row-stochastic matrices. Following (Kozitsin, 2022), this matrix is called the *transition matrix*. This matrix encodes how social influence processes unfold on the social network. In particular, using this matrix, one could approach different micro influence mechanisms, including assimilative influence, bounded confidence, and dissimilative influence, as well as their various mathematical forms (linear, inverted U-shaped). Further, this matrix (which is actually the key element of the model and stores all the information regarding the organization of social influence processes) can be easily calibrated on real (necessarily longitudinal) data by just following the classical definition of probability. These properties make the model pretend to be a unified framework that could bring together theoretical and practical sides of the opinion formation models theory.

In the rest of the paper, this model will be referred to as the Basic Model.

### 3.2. Mean-field approximation for the Basic Model

The mean-field approximation for the Basic Model was elaborated in (Kozitsin, 2022) under assumptions that (i) the number of agents in the system is sufficiently large and (ii) the social network is a complete graph (so-called fully mixed approximation). As a result, the following system of ordinary differential equations can be obtained:

$$\frac{dy_i(\tau)}{d\tau} = \sum_{s,l,k} y_s(\tau) y_l(\tau) p_{s,l,k} \big( \delta_{k,i} - \delta_{s,i} \big),$$

or simply

$$\frac{dy_i(\tau)}{d\tau} = \sum_{s,l} y_s(\tau) y_l(\tau) p_{s,l,i} - y_i(\tau), \tag{1}$$

where $i \in [m]$ and $\delta_{i,j}$ is the Kronecker delta. In system (1), quantities $y_i(\tau)$ measure the fractions of opinion camps $x_i$ and are defined as

$$y_i(\tau) = \frac{Y_i(\tau)}{N},$$

where $Y_i(\tau)$ is the number of agents espousing opinion $x_i$ at the scaled time moment $\tau$ ($\tau = \frac{t}{N}$ and $\delta\tau = \frac{1}{N}$):

$$Y_i(\tau) = \#\{ j \mid o_j(\tau) = x_i \}.$$

### 3.3. Introducing stubborn agents

The natural extension of the Basic Model is considering agents with different levels of persuasiveness and susceptibility to influence. In its canonical version, the Basic Model assumes that agents with similar opinions respond (on average) equally to the same stimuli – just because they all are described by the same transition matrix that encodes individuals' behaviour as a function of their opinions. In fact, this is not a realistic representation of the real world whereby individuals' behaviour may vary across many different factors, not only their views. For example, empirical studies suggest that younger individuals tend to be more susceptible to social influence than elder people. Further, some research has revealed that women socialize better than men (Peshkovskaya *et al.*, 2018, 2019). The general approach to handling such situations was briefly discussed in (Kozitsin, 2022) and is drawn upon the idea of using several transition matrices over the whole population at once. Here, the current paper continues this idea by elaborating on the model that considers two types of agents. Assume that their populations do not change over time and are denoted by $N^y$ and $N^u$, where $N^y + N^u = N$. Accordingly, $n^y = N^y/N$ and $n^u = N^u/N$. The first-type agents can be understood as, for example, ordinary (native) users of online networking services. Influence events that arise between such agents are encoded by the transition matrix $P^y$. Second-type agents are assumed to be unsusceptible to influence and thus do not change their opinions whatever happens. However, they could influence the first-type agents, and this cross-type influence is depicted by the transition matrix $P^u$. One could think about the second-type agents as online bots. Because these agents do not change their opinions, the matrices $P^y$ and $P^u$ describe exhaustively the map of influence events in the system. For the sake of clarity, in what follows, the first-type agents will be referred to as *ordinary* agents. The second-type agents will be denoted as *stubborn* agents.

Suppose that the social network $G$, for now, includes both ordinary and stubborn agents. For the time being, their location is not specified, but in real examples, one can assume that the stubborn agents are located on the periphery of the network – as do online bots (González-Bailón & De Domenico, 2021). The dynamics of the new model go as follows. At each time moment $t$, a randomly chosen agent $i$ (no matter, ordinary or stubborn) is influenced by one of its neighbours – say, agent $j$ – who is also chosen by chance. Of course, if the stubborn agent is selected, then nothing happens. Otherwise, in the case of choosing an ordinary agent, its opinion $o_i(t)$ changes while being influenced by $o_j(t)$. Depending on the type of agent $j$, the outcome of influence is determined by the distribution $\{p^y_{s,l,1}, \dots, p^y_{s,l,m}\}$ (if $j$ is an ordinary agent) or $\{p^u_{s,l,1}, \dots, p^u_{s,l,m}\}$ (if $j$ is a stubborn agent).

This approach is quite flexible because it assumes that ordinary agents may perceive in-type and cross-type influence differently. This contrast may be explained by the following intuition. Communications that native users have with online bots tend to have an accidental character (random meetings in online conversations), whereas influence from online friends/followees tends to spread over ties established and approved by users themselves. As such, one should expect that influence coming from online friends (native users) is more solid than those from bots. However, there could be other, completely inverse, ideas. For example, online bots are accused of disseminating fake content that is extremely viral and persuasive, whereas information received from close (online) friends, according to Granovetter's famous theory of weak ties, is hardly so shocking and convincing ( Granovetter, 1977).

The resulting model is called the Advanced Model below.

### 3.4. Mean-field approximation for the Advanced Model

For now, the opinion distribution of the system at time moment $\tau$ is presented by the vector:

$$\begin{bmatrix} \vec{y}(\tau) \\ \vec{u}(\tau) \end{bmatrix} \in \mathbb{R}^{2m},$$

where

$$\vec{y}(\tau) = \begin{bmatrix} y_1(\tau) \\ ... \\ y_m(\tau) \end{bmatrix} \in \mathbb{R}^m, \vec{u}(\tau) = \begin{bmatrix} u_1(\tau) \\ ... \\ u_m(\tau) \end{bmatrix} \in \mathbb{R}^m,$$

and the quantities $y_i$ and $u_i$ store the fraction of ordinary and stubborn agents that espouse opinion $x_i$. Because stubborn agents do not change their opinions, the dynamics of ordinary agents' opinions are of particular interest. Below, it is assumed that the stubborn agents' opinions are defined from the outside. Technically, it means that the function $\vec{u}(\tau)$ is exogenous to the model. Further, $\vec{u}(\tau)$ should satisfy:

$$\vec{u}(\tau) \geq \vec{0}, u_1(\tau) + \cdots + u_m(\tau) = n^u.$$

For the Advanced model, the following mean-field approximation can be obtained (see Appendix A for derivations):

$$\frac{d\vec{y}(\tau)}{d\tau} = \vec{f}\big(\vec{y}(\tau), \vec{u}(\tau)\big), \tag{2}$$

where $\vec{f} = [f_1 \quad \dots \quad f_m]^T$ and

$$f_i = \sum_{s,l,k} y_s(\tau)\big[y_l(\tau)p_{s,l,k}^y + u_l(\tau)p_{s,l,k}^u\big](\delta_{k,i} - \delta_{s,i}). \tag{3}$$

**Remark 1.** *Note that to obtain this mean-field approximation, one should assume that the exogenous function $\vec{u}(\tau)$ changes "no faster" than $\vec{y}(\tau)$. Otherwise, the law of large numbers that necessarily appears in derivations cannot be applied.*

In (3), the variables $p_{s,l,i}^y$ and $p_{s,l,i}^u$ specify all probabilities of opinion shifts that end up in opinion $x_i$. A straightforward observation from system (2) is that one of its first integrals is

$$y_1 + \dots + y_m = n^a.$$

To prove this property, one needs only to notice that

$$\sum_i f_i = \sum_i \sum_{s,l,k} y_s(\tau)\big[y_l(\tau)p_{s,l,k}^y + u_l(\tau)p_{s,l,k}^u\big](\delta_{k,i} - \delta_{s,i}) = 0.$$

As such, expression (3) can be rewritten as follows:

$$f_i = \sum_{s,l} y_s(\tau)\big[y_l(\tau)p_{s,l,k}^y + u_l(\tau)p_{s,l,k}^u\big] - y_i(\tau)\left(\sum_j y_j + \sum_j u_j\right)$$

$$= \sum_{s,l} y_s(\tau)\big[y_l(\tau)p_{s,l,k}^y + u_l(\tau)p_{s,l,k}^u\big] - y_i(\tau).$$

**Remark 2.** *While obtaining the last equality, it has been silently assumed that each stubborn agent is speechless and always translates some opinion. Otherwise, if some stubborn agents "pass their turns", then one should write (see Subsection 4.5, where this special situation is the primary focus):*

$$f_i = \sum_{s,l} y_s(\tau)\big[y_l(\tau)p_{s,l,k}^y + u_l(\tau)p_{s,l,k}^u\big] - y_i(\tau)\left(n^y + \sum_j u_j\right),$$

*where $\sum_j u_j$ constitutes the fraction of stubborn agents that are not silent.*

Equation (2) can be advanced with the initial condition

$$\vec{y}(t_0) = \vec{y}^0 = [y_1^0 \quad \dots \quad y_m^0]^T, \qquad (4)$$

where $y_1^0 + \cdots + y_m^0 = n^y$, and $\vec{y}^0 \geq \vec{0}$. The resulted Cauchy problem satisfied the following statement.

**Statement 1** *(See Appendix B for the proof). The Cauchy problem (2), (4) has a unique solution $\vec{y}(\tau)$, which is a nonnegative function with all the components amounting to $n^y$. This solution is an analytic function of the models' parameter $\vec{u}(\tau)$ and the initial condition $\vec{y}^0$. The solution can be extended to the whole axis.*

## 4. Control problem

### 4.1. Problem formulation

Let us assume that a concerned person (politician, commercial company, blogger – hereafter the Person) tries to influence the social system by aiming to approach a desirable opinion distribution (e.g., to convince citizens to be vaccinated or to lead a healthy lifestyle). Assume that the Person can approach their goal only by employing stubborn agents for their needs. It will be also supposed that there are a fixed number of stubborn agents in the system, and the Person should adjust their opinions. For simplicity, it is assumed that the Person should employ all the stubborn agents existing in the system. It means that at every moment all the stubborn agents' opinions should be defined (there should be no silent stubborn agents that stand idle). In Subsection 4.5, possible extensions of this formulation will be discussed.

Assume that the number of agents (of all types) is high, and all possible communications between agents can occur (on an equal basis). As such, the mean-field approximation (2) is a plausible description of the system. The system starts from time moment $\tau = t_0$ with a fixed opinion distribution $\vec{y}^0 \in \mathbb{R}^m$. The purpose of the Person is to choose the function $\vec{u}(\tau)$ which minimizes an objective functional $J$ that represents the Person's desirable opinion distribution:

$$\begin{aligned} & J \to \inf_{\vec{u}(\tau)}, \\ & s.t. \, \vec{u}(\tau) \in U, \\ & \frac{d\vec{y}(\tau)}{d\tau} = \vec{f}\big(\vec{y}(\tau), \vec{u}(\tau)\big), \\ & \vec{y}(t_0) = \vec{y}^0. \end{aligned} \tag{5}$$

The control set $U$ includes all possible admissible control functions. More precisely, it includes all piecewise continuous functions such that

$$u_1(\tau) + \cdots + u_m(\tau) = n^u, \tag{R1}$$

$$\vec{0} \leq \vec{u}(\tau). \tag{R2}$$

Let us consider the following family of cost functionals:

$$J = \int_{t_0}^{t_1} g\big(\vec{w}, \vec{y}(\tau)\big) d\tau + G\big(\vec{v}, \vec{y}(t_1)\big), \tag{6}$$

where $g\big(\vec{w}, \vec{y}(\tau)\big) = w_1 y_1(\tau) + \cdots + w_m y_m(\tau)$, $G\big(\vec{v}, \vec{y}(t_1)\big) = v_1 y_1(t_1) + \cdots + v_m y_m(t_1)$, and the terminal time $t_1$ is fixed. *Opinion-weight* vectors $\vec{w}$ and $\vec{v}$ are assumed to be predetermined. For convenience, it is supposed that these vectors are nonnegative. Within this encoding strategy, the magnitudes of components of these vectors represent those opinion camps that should be minimized foremost. For example, the vector $[1 \quad 0 \quad 0]^T$ (that belongs to the triple opinion space $m = 3$) indicates that the Person aims at minimizing the number of agents with opinion $x_1$. In turn, the vector $[2 \quad 1 \quad 0]^T$ minimizes those factions $y_1$ and $y_2$ that represent the left-sided opinions $x_1$ and neutral ones $x_2$ while giving special attention to individuals espousing the left position. In fact, the assumption of nonnegative opinion-weight vectors does not constrain the Person at all: to maximize the number of a certain opinion camp, the Person should mitigate the populations of other camps. As such, the Person has no need for negative opinion weights.

Because control problem (5) is affine in control functions and for each $\tau$ the control set is compact and convex, then, drawing upon the Filippov theorem, one can guarantee that the reachable set is compact. As such, the following statement is true.

***Statement 2.*** *An optimal control in control problem (5) exists.*

***Remark 3.*** *In the general case, one cannot guarantee that control problem (5) has a unique solution. Below (see Subsection 5.3, Scenario 1), specific examples of this issue will be provided. In a nutshell, the general idea behind this is that for long time intervals, if the system is "well-controlled" (in the sense it can be pushed both to the left and right endpoints of the opinion spectrum in finite time regardless of its initial position) and the cost functional includes only the terminal term, then it could be more than one optimal trajectory leading to a desirable opinion distribution.*

## 4.2. Pontryagin Maximum Principle

The Pontryagin maximum principle formulates a necessary condition for optimality in terms of the Hamiltonian function (Lenhart & Workman, 2007). Let us apply this principle to the optimal control problem (5) with cost functional (6). First, write the Hamiltonian function:

$$H\left(\tau, \vec{y}(\tau), \vec{u}(\tau), \vec{\lambda}(\tau)\right) = \sum_i \lambda_i(\tau) f_i\left(\vec{y}(\tau), \vec{u}(\tau)\right) - g\left(\vec{y}(\tau)\right), \tag{7}$$

where $\vec{\lambda} = [\lambda_1 \quad ... \quad \lambda_m]^T$ is the adjoint function. According to the Pontryagin maximum principle, if $\vec{u}^*(\tau)$ is the optimal control and $\vec{y}^*(\tau)$ is the corresponding trajectory of state variables, then there exists such a (piecewise differentiable) adjoint function $\vec{\lambda}(\tau)$ that satisfies the following system of differential equations:

$$\frac{d\lambda_i}{d\tau} = -\frac{\partial H\left(\tau, \vec{y}^*(\tau), \vec{u}^*(\tau), \vec{\lambda}(\tau)\right)}{\partial y_i},$$
$$\lambda_i(t_1) = -\frac{\partial G}{\partial y_i}. \tag{8}$$

Further, for each admissible control $\vec{u}(\tau)$, one can obtain

$$H\left(\tau, \vec{y}^*(\tau), \vec{u}^*(\tau), \vec{\lambda}(\tau)\right) = \max_{\vec{u}(\tau) \in U} H\left(\tau, \vec{y}^*(\tau), \vec{u}(\tau), \vec{\lambda}(\tau)\right) \tag{9}$$

at each time moment $\tau$.

The Euler-Lagrange equations (8) may be presented as follows:

$$\frac{d\vec{\lambda}(\tau)}{d\tau} = A\vec{\lambda}(\tau) + \vec{w},$$
$$\vec{\lambda}(t_1) = -\vec{v}. \tag{10}$$

In (10), $\vec{w}$ and $\vec{v}$ are opinion-weight vectors inherited from the cost functional, $A$ is an $m \times m$ matrix whose components $a_{ij}(\vec{y}^*, \vec{u}^*)$ are defined as follows (see Appendix C for technical details):

$$a_{ij}(\vec{y}, \vec{u}) = \delta_{i,j} - \sum_s y_s p_{s,i,j}^y - \sum_l (y_l + u_l)p_{i,l,j}^u.$$

**Statement 3.** *(See Appendix D for proof). Let $\vec{v} \geq \vec{0}$ and $\vec{w} \geq \vec{0}$. Let $\vec{\lambda}$ be the solution to the Cauchy problem (10) on the interval $[t_0, t_1]$ for some control and state functions $\vec{u}$ and $\vec{y}$ (not necessarily optimal). Then the solution $\vec{\lambda}(\tau)$ of the Cauchy problem (10) as well as all the partial derivatives $\frac{\partial H}{\partial u_i}$ are nonpositive.*

**Remark 4.** *In fact, the nonpositiveness of partial derivatives $\frac{\partial H}{\partial u_i}$ should not be a surprise here. If one forgets about restriction (R1), then the guaranteed way to minimize the cost functional would be to choose $\vec{u}(\tau) \equiv \vec{0}$ because in this case, all the state variables will tend to zero. To prove this property, let us take $\vec{u}(\tau) \equiv \vec{0}$ and consider the summation of all right parts (3):*

$$\sum_i f_i = \sum_i \left\{ \sum_{s,l} y_s \left[ y_l p_{s,l,i}^y + \underbrace{u_l}_{=0} p_{s,l,i}^u \right] - y_i \right\} = \sum_{s,l} y_s y_l \underbrace{\sum_i p_{s,l,i}^y}_{1} - \sum_i y_i = \sum_{s,l} y_s y_l - n^y$$

$$= n^y(n^y - 1) < 0.$$

*Because $\sum_i f_i$ is negative, then the sum of all state variables should decrease. The state variables are nonnegative (proof of this fact follows the same line of reasoning as that for Statement 1), and thus all they should converge to zero.*

Let us now take a closer look at the organization of the optimal control. The structure of the control set $U$ and the linearity of the Hamiltonian with respect to $\vec{u}$ ensure the following statement.

**Statement 4.** *Let $\frac{\partial H(\tau)}{\partial u_i} > \frac{\partial H(\tau)}{\partial u_j}$ on interval $I_\tau$ for some $i \in [m]$ and all $j \in [m]/\{i\}$ on the optimal trajectory $\vec{y}^*, \vec{u}^*$. Then the optimal control $\vec{u}^*$ should satisfy*

$$\vec{u}^*(\tau) = \begin{bmatrix} 0 & \cdots & 0 & \underbrace{n^b}_{i-\text{th component}} & 0 & \cdots & 0 \end{bmatrix}^T$$

*for $\tau \in I_\tau$.*

The problem, however, arises if there are several partial derivatives $\frac{\partial H(\tau)}{\partial u_i}$ with the highest value at once (singular control). In this situation, more complex techniques should be applied.

**Statement 5.** *(See Appendix E for proof). Assume that there are $d$ partial derivatives $\frac{\partial H(\tau)}{\partial u_{i_1}}, \ldots, \frac{\partial H(\tau)}{\partial u_{i_d}}$ with the highest value on the optimal trajectory $\vec{y}^*, \vec{u}^*$ for $\tau \in I_\tau$. In this case, the components of the optimal control $\vec{u}^*$ are zero for $i \notin \{i_1, \ldots, i_d\}$ on interval $I_\tau$. The rest ones $(u_{i_1}^*, \ldots, u_{i_d}^*)$ satisfy the following system:*

$$\begin{cases} C^{i_1 i_2} + c_{i_1}^{i_1 i_2} u_{i_1}^* + \cdots + c_{i_d}^{i_1 i_2} u_{i_d}^* = 0, \\ C^{i_2 i_3} + c_{i_1}^{i_2 i_3} u_{i_1}^* + \cdots + c_{i_d}^{i_2 i_3} u_{i_d}^* = 0, \\ \qquad\qquad \cdots \\ C^{i_{d-1} i_d} + c_{i_1}^{i_{d-1} i_d} u_{i_1}^* + \cdots + c_{i_d}^{i_{d-1} i_d} u_{i_d}^* = 0, \end{cases} \tag{11}$$

*on this interval, where*

$$C^{ij} = \sum_{r,q} \left( y_q^* \left\{ w_r - \sum_k \lambda_k \left[ \sum_s y_s^* p_{s,r,k}^y + \sum_l y_l^* p_{r,l,k}^u - \delta_{r,k} \right] \right\} \right.$$
$$\left. + \lambda_r \left\{ \sum_{s,l} y_s^* y_l^* p_{s,l,q}^y - y_q^* \right\} \right) \left( p_{q,i,r}^u - p_{q,j,r}^u \right)$$

*and*

$$c_i^{ij} = \sum_{r,q,k} \left[ \lambda_r y_k^* p_{k,i,q}^u - y_q^* \lambda_k p_{r,i,k}^u \right] \left( p_{q,i,r}^u - p_{q,j,r}^u \right), c_j^{ij}$$

$$= \sum_{r,q,k} \left[ \lambda_r y_k^* p_{k,j,q}^u - y_q^* \lambda_k p_{r,j,k}^u \right] \left( p_{q,i,r}^u - p_{q,j,r}^u \right).$$

**Remark 5.** *Recall that in system (11), variables $u_{i_1}^*, \ldots, u_{i_d}^*$ should also satisfy $u_{i_1}^* + \cdots + u_{i_d}^* = n^u$. As such, one of variables in (11) is redundant, and one ends up with the linear system of $m - 1$ equations with respect to $m - 1$ variables (say, with respect to $u_{i_1}^*, \ldots, u_{i_{d-1}}^*$):*

$$\begin{bmatrix} c_{i_1}^{i_1 i_2} - c_{i_d}^{i_1 i_2} & \cdots & c_{i_{d-1}}^{i_1 i_2} - c_{i_d}^{i_1 i_2} \\ \vdots & \ddots & \vdots \\ c_{i_1}^{i_{d-1} i_d} - c_{i_d}^{i_{d-1} i_d} & \cdots & c_{i_{d-1}}^{i_{d-1} i_d} - c_{i_d}^{i_{d-1} i_d} \end{bmatrix} \begin{bmatrix} u_{i_1}^* \\ \vdots \\ u_{i_{d-1}}^* \end{bmatrix} = \begin{bmatrix} -C^{i_1 i_2} - c_{i_d}^{i_1 i_2} \\ \vdots \\ -C^{i_{d-1} i_d} - c_{i_d}^{i_{d-1} i_d} \end{bmatrix}. \tag{12}$$

### 4.3. Forward-backward sweep method

Problem (5) is hardly solvable analytically, so let us try to solve it numerically. Preliminary experiments revealed that direct methods featured low efficiency and needed a huge amount of time to proceed. In contrast, the indirect forward-backward sweep (FBS) method demonstrated better results. Below, the FBS method will be used as a workhorse in numerical experiments. Starting from some initial guess $\vec{u}^0(\tau)$, this numerical algorithm seeks an optimal control $\vec{u}^*$ drawing upon the Pontryagin maximum principle. The FBS method derives a control function on a grid of predefined points $Gr = \{t_0 = \tau_0, \tau_1, \ldots, \tau_{n-1}, \tau_n = t_1\}$ by iterating the following steps:

**Step 1.** Initialize $\vec{u}(\tau)$. If it is the first iteration, then take the initial guess $\vec{u}(\tau) := \vec{u}^0(\tau)$. Otherwise, apply $\vec{u}(\tau) := \vec{u}_{new}(\tau)$ from **Step 4**.

**Step 2.** For a given initial point $\vec{y}^0$ and the control function $\vec{u}(\tau)$ from **Step 1**, calculate $\vec{y}(\tau)$ by solving the Cauchy problem (2), (4).

**Step 3.** Substitute the values for $\vec{u}(\tau)$ and $\vec{y}(\tau)$ into the Cauchy problem (10) and solve it (with respect to $\vec{\lambda}(\tau)$).

**Step 4.** Building upon (9) and using $\vec{y}(\tau)$ and $\vec{\lambda}(\tau)$ that were found in previous steps, maximize the Hamiltonian pointwise with respect to $\vec{u}(\tau)$. By doing so, find some new approximation $\vec{u}_{new}(\tau)$.

**Step 5.** Check whether $\vec{u}_{new}(\tau)$ differs from $\vec{u}(\tau)$. If there is no difference (or, the difference is small), then stop the algorithm and give the output $\vec{u}_{new}(\tau)$. Otherwise, start a new iteration by returning to **Step 1**.

Technical details on how the FBS method is implemented in computations can be found in Appendix F. Below, the terms "FBS method" and "algorithm" will be used interchangeably. The summary of the algorithm arguments is presented in Table 1.

Table 1. The arguments of the algorithm and their definitions

| Argument | Definition |
|---|---|
| $m$ | Dimensionality of the opinion space. |
| $P^y, P^u \in \mathbb{R}^{m \times m \times m}$ | Transition matrices that describe influence events between ordinary agents ($P^y$) and those that appear when stubborn agents influence ordinary ones ($P^u$). |
| $n^y, n^u$ | Fractions of ordinary and stubborn agents ($n^y + n^u = 1, n^y \geq 0, n^u \geq 0$). |
| $\vec{w}, \vec{v} \in \mathbb{R}^m$ | Opinion-weight vectors that define the integral and terminal parts of the cost functional (see (6)) |
| $\vec{y}^0 \in \mathbb{R}^m$ | The initial opinion distribution of ordinary agents (at time $t_0$) |
| $\vec{u}^0(\tau) \in \mathbb{R}^m$ | Initial guess for the control |
| $Gr = \{\tau_0, \tau_1, \dots, \tau_{n-1}, \tau_n\}$ | The grid of points on which the control should be approximated ($\tau_0 = t_0, \tau_n = t_1$) |

### 4.4. Drawbacks of the FBS method

Before demonstrating how the FBS method operates through concrete examples, let us first discuss its possible drawbacks. First and foremost, it builds upon the maximum principle, which is a necessary condition, not a sufficient one. In this regard, even if the pair $\vec{u}(\tau)$ and $\vec{y}(\tau)$ satisfies conditions (8)–(9), it does not necessarily follow that this pair is optimal. Next, one iteration of the algorithm casts two auxiliary numerical methods (which solve the Cauchy problems in **Steps 2** and **3** – see Appendix F for details). Each of them approximates the corresponding solutions with some errors (especially when integrating long time intervals), and when it comes to maximizing the Hamiltonian (in **Step 4**), the resulting error may be quite huge.

Further, if one approximates the solution to the Cauchy problem (2), (4), restriction (R1) may no longer work in some cases (in a typical situation, one can observe that all state variables tend to zero: $y_i \to 0$; besides, it could be the case that the state variables grow unrestrictedly) – such issues arise for long time intervals and for certain configurations of transition matrices (see Section 5, Scenario 2). To some extent, this problem is fixed by adding some modifications to the Runge-Kutta method (see Appendix F for details); however, even after implementing these modifications, some transition matrices still lead to difficulties even in a relatively short time intervals.

One more drawback of the FBS method is that in some cases (see Subsection 5.3), its outputs depend on what initial guess is applied. In this regard, using only one guess typically leads to low efficiency.

### 4.5. Possible generalizations of the control problem

Let us present some generalizations of the control problem (5) which can also be of some interest.

First, one could assume that not all the stubborn agents should disseminate opinions at a particular time moment – it could be a case that the better strategy is silence. To model this situation, one could consider the following control problem:

$$J \to \inf_{\vec{u}(\tau)},$$

$$s.t. \, \vec{u}(\tau) \in U,$$

$$\frac{dy_i}{d\tau} = \sum_{s,l} y_s(\tau) \left[ y_l(\tau) p_{s,l,i}^y + u_l(\tau) p_{s,l,i}^u \right] - y_i(\tau) \left( n^y + \underbrace{u_1 + \cdots + u_m}_{\leq n^u} \right), i \in [m],$$

$$\vec{y}(t_0) = \vec{y}^0,$$

where the control set has the following structure now:

$$U = \{ u_1(\tau) + \cdots + u_m(\tau) \leq n^u, 0 \leq u_i(\tau) \leq n^u \}.$$

The second generalization concerns the situation when the Person can control only some fraction of all stubborn agents and the opinions of other ones are defined by someone else (by an opponent). More precisely, if the Person has control over $n_P^u \leq n^u$ of all stubborn agents, then one ends up with the following problem (for simplicity, it is again assumed that each stubborn agent is not silent and translates an opinion at each time moment):

$$J \to \inf_{\vec{u}^P(\tau)},$$

$$s.t. \, \vec{u}_P(\tau) \in U,$$

$$\frac{dy_i}{d\tau} = \sum_{s,l} y_s(\tau) y_l(\tau) p_{s,l,i}^y + \sum_s y_s(\tau) \sum_{l \in I^b / I_P^b} u_l(\tau) p_{s,l,i}^u + \sum_s y_s(\tau) \sum_{l \in I_P^b} u_l(\tau) p_{s,l,i}^u - y_i(\tau), i \in [m],$$

$$\vec{y}(t_0) = \vec{y}^0,$$

where $I^b$ and $I_P^b$ represent indices of all the stubborn agents and only those that are controlled by the Person correspondingly. The function $\vec{u}^P(\tau)$ stands for the opinions of the Person's stubborn agents. The more interesting would be to consider that there are two concerned persons who want to influence the opinions of ordinary agents via the stubborn ones. In this particular situation, one should use the methods from the special branch of the game theory named differential games.

The rest of the paper will focus on the control problem (5). Generalizations that were presented in this Subsection are beyond the scope of this article and will be covered in ongoing studies.

## 5. Case studies

This Section demonstrates how the FBS method works through concrete examples. Three basic Scenarios are considered. In all of them, agents change their opinions in the five-fold opinion space $m = 5$. This dimensionality is sufficiently small to ensure the possibility of interpretations and, at the same time, gives the opportunity to consider main micro-foundations regarding social influence (Flache *et al*., 2017). More precisely, Scenario 1 analyses the assimilative influence mechanism, Scenario 2 covers the case of the bounded confidence mechanism, and Scenario 3 concerns the coexistence of assimilative and dissimilative influence mechanisms. The flexibility of the Advanced Model provides the perfect opportunity to cover all these Scenarios within the same theoretical framework by using different transition matrices.

To implement these three mechanisms, it is necessarily assumed that opinions $x_1, x_2, x_3, x_4, x_5$ are arranged:

$$x_1 \prec x_2 \prec x_3 \prec x_4 \prec x_5.$$

Just for interpretation purposes, let us suppose that these opinion values are specified with numbers:

$$x_1 = -2, x_2 = -1, x_3 = 0, x_4 = 1, x_5 = 2.$$

For now, it is possible to introduce the distance in the opinion space, which is measured as the absolute difference in opinion values, just as follows: $|x_i - x_j| = |i - j|$.

For simplicity, in each Scenario, it is also assumed that $P^y = P^u = P$, that is, ordinary agents perceive influence coming from the same agents and from stubborn ones on an equal basis. Let us now discuss these Scenarios in more detail.

### 5.1. Scenarios

*Scenario 1. Assimilative social influence*

This Scenario analyses the situation when opinion dynamics are governed by the assimilative social influence assumption, whereby interactions between agents reduce the absolute difference in opinions and the greater initial opinion discrepancies lead to more distant opinion shifts (that is, more distant opinions are more attractive). The classic example of this assumption can be found in the famous DeGroot model (DeGroot, 1974), in which agents follow the *linear* assimilative social influence assumption. Here, a stylized version of the assimilative assumption is considered by employing the following (manually created) transition matrix:

$$P_1 = \begin{bmatrix} 1 & 0 & 0 & 0 & 0 \\ .97 & .03 & 0 & 0 & 0 \\ .94 & .04 & .02 & 0 & 0 \\ .91 & .05 & .03 & .01 & 0 \\ .87 & .06 & .04 & .02 & .01 \end{bmatrix}, P_2 = \begin{bmatrix} .03 & .97 & 0 & 0 & 0 \\ 0 & 1 & 0 & 0 & 0 \\ 0 & .97 & .03 & 0 & 0 \\ 0 & .94 & .04 & .02 & 0 \\ 0 & .91 & .05 & .03 & .01 \end{bmatrix},$$

$$P_3 = \begin{bmatrix} .02 & .04 & .94 & 0 & 0 \\ 0 & .03 & .97 & 0 & 0 \\ 0 & 0 & 1 & 0 & 0 \\ 0 & 0 & .97 & .03 & 0 \\ 0 & 0 & .94 & .04 & .02 \end{bmatrix}, P_4 = \begin{bmatrix} .01 & .03 & .05 & .91 & 0 \\ 0 & .02 & .04 & .94 & 0 \\ 0 & 0 & .03 & .97 & 0 \\ 0 & 0 & 0 & 1 & 0 \\ 0 & 0 & 0 & .97 & .03 \end{bmatrix}, \quad (S1)$$

$$P_5 = \begin{bmatrix} .01 & .02 & .04 & .06 & .87 \\ 0 & .01 & .03 & .05 & .91 \\ 0 & 0 & .02 & .04 & .94 \\ 0 & 0 & 0 & .03 & .97 \\ 0 & 0 & 0 & 0 & 1 \end{bmatrix}.$$

The transition matrix (S1) ensures only the assimilative opinion shifts and indicates that a higher difference in opinions between interacting agents increases the probability of opinion change. Further, the more the difference, the higher the expectable magnitude of the opinion shift. In line with empirical studies (Kozitsin, 2021; Moussaïd *et al.*, 2013), I rely on the assumption of high opinion resistance towards social influence: even if the opposite opinions $x_1$ and $x_m$ communicate, then the probability of opinion change is only 0.13.

It is straightforward to note that within the opinion dynamics governed by the transition matrix (S1), if the Person focuses on pushing the system towards, say, the left endpoint of the opinion spectrum, then he/she should apply the stubborn agents disseminating the most radical left-side opinion:

$$\vec{u}(\tau) \equiv [.4 \quad 0 \quad 0 \quad 0 \quad 0]^T \tag{13}$$

or, simply, influence the system with opinion $x_1$.

*Scenario 2. Bounded confidence*

Within the bounded-confidence assumption, assimilative influence may occur only if the opinions of interacting agents are not too distant – otherwise, the recipient rejects the influence and remains on the same opinion. To encode this situation, the following transition matrix was first used:

$$P_1 = \begin{bmatrix} 1 & 0 & 0 & 0 & 0 \\ .9 & .1 & 0 & 0 & 0 \\ 1 & 0 & 0 & 0 & 0 \\ 1 & 0 & 0 & 0 & 0 \\ 1 & 0 & 0 & 0 & 0 \end{bmatrix}, P_2 = \begin{bmatrix} .9 & 1 & 0 & 0 & 0 \\ 0 & 1 & 0 & 0 & 0 \\ 0 & .9 & .1 & 0 & 0 \\ 0 & 1 & 0 & 0 & 0 \\ 0 & 1 & 0 & 0 & 0 \end{bmatrix},$$

$$P_3 = \begin{bmatrix} 0 & 0 & 1 & 0 & 0 \\ 0 & .1 & .9 & 0 & 0 \\ 0 & 0 & 1 & 0 & 0 \\ 0 & 0 & .9 & .1 & 0 \\ 0 & 0 & 1 & 0 & 0 \end{bmatrix}, P_4 = \begin{bmatrix} 0 & 0 & 0 & 1 & 0 \\ 0 & 0 & 0 & 1 & 0 \\ 0 & 0 & .1 & .9 & 0 \\ 0 & 0 & 0 & 1 & 0 \\ 0 & 0 & 0 & .9 & .1 \end{bmatrix}, \tag{S2}$$

$$P_5 = \begin{bmatrix} 0 & 0 & 0 & 0 & 1 \\ 0 & 0 & 0 & 0 & 1 \\ 0 & 0 & 0 & 0 & 1 \\ 0 & 0 & 0 & .1 & .9 \\ 0 & 0 & 0 & 0 & 1 \end{bmatrix}.$$

According to the transition matrix (S2), only neighbouring opinions can communicate and the probability of adoption in such communications is 0.1. However, for the transition matrix (S2), the FBS method experiences difficulties in obtaining the solution to the Cauchy problem (2), (4). Importantly, even the modification of the Runge-Kutta algorithm fails to derive a valid solution

(see Appendix F for details). To avoid this roadblock, let us consider a slightly modified version

of the transition matrix (S2), which is free from this problem:

$$P_1 = \begin{bmatrix} 1 & 0 & 0 & 0 & 0 \\ .9 & .1 & 0 & 0 & 0 \\ .92 & .06 & .02 & 0 & 0 \\ .94 & .04 & .015 & .005 & 0 \\ .96 & .03 & .005 & .003 & .002 \end{bmatrix}, P_2 = \begin{bmatrix} .9 & .1 & 0 & 0 & 0 \\ 0 & 1 & 0 & 0 & 0 \\ 0 & .9 & .1 & 0 & 0 \\ 0 & .92 & .06 & .02 & 0 \\ 0 & .94 & .04 & .015 & .005 \end{bmatrix},$$

$$P_3 = \begin{bmatrix} .02 & .06 & .92 & 0 & 0 \\ 0 & .1 & .9 & 0 & 0 \\ 0 & 0 & 1 & 0 & 0 \\ 0 & 0 & .9 & .1 & 0 \\ 0 & 0 & .92 & .06 & .02 \end{bmatrix}, P_4 = \begin{bmatrix} .005 & .015 & .04 & .94 & 0 \\ 0 & .02 & .06 & .92 & 0 \\ 0 & 0 & .1 & .9 & 0 \\ 0 & 0 & 0 & 1 & 0 \\ 0 & 0 & 0 & .9 & .1 \end{bmatrix}, \quad (S2')$$

$$P_5 = \begin{bmatrix} .002 & .003 & .005 & .03 & .96 \\ 0 & .005 & .015 & .04 & .94 \\ 0 & 0 & .02 & .06 & .92 \\ 0 & 0 & 0 & .1 & .9 \\ 0 & 0 & 0 & 0 & 1 \end{bmatrix}.$$

Within the transition matrix (S2'), the probability of opinion change is positive if and only

if the influence opinion $x_j$ differs from those of the recipient $x_i$ and takes the maximal value if $x_j$

and $x_i$ are neighbouring opinions. If the distance between these opinions increases, then the probability of opinion change goes down: for opposite positions $x_1$ and $x_5$, this probability is equal to 0.04, whereas for neighbouring ones, the same quantity is 0.1. Further, the expected magnitude of opinion change is also negatively associated with $|x_i - x_j|$: if $|x_i - x_j| = 4$, then the expected magnitude is 0.057, if $|x_i - x_j| = 2$, then one obtains 0.08, for $|x_i - x_j| = 1$ the excepted opinion change magnitude is 0.1. In this regard, the transition matrix (S2') may be considered as a different,

smoother, representation of the bounded confidence assumption. Further, this approach is likely

more realistic than the strictly bounded confidence assumption implemented in (S2) (Kurahashi-

Nakamura *et al.*, 2016).

This situation is more interesting from the perspective of solving the control problem,

compared to those in the first Scenario. Within the bounded confidence assumptions, influencing the system with a radical opinion without accounting for the agents' opinions is likely not the best strategy: perhaps, the Person should find some middle ground by toning down his/her arguments.

Consider the following stylized situation. Let us assume that the system is located on the right endpoint of the opinions spectrum ($x_5$) and the Person strives to push agents' opinions towards the left endpoint ($x_1$). In this case, the optimal control strategy is quite obvious (at least, from a qualitative perspective). The Person should apply a "nudging" strategy whereby stubborn agents start with disseminating the closest (and thus the most profitable) opinion $x_4$ and then gradually slide their positions towards the left (use opinion $x_3$ instead of $x_4$ from some moment, then opinion $x_2$ and so on) as the system moves in the same direction.

*Scenario 3. Empirical data (coexistence of assimilative and dissimilative opinion shifts)*

Within the third Scenario, a transition matrix calibrated on empirical data from (Kozitsin, 2020, 2021) is employed. These data represent the opinion dynamics of a large-scale sample (~1.6 M) of online social network users around a political topic. As was reported in (Kozitsin, 2021), within these data, one can observe a quite complex asymmetric map of influence processes, with both assimilative and dissimilative (directed opposite to the influence source) opinion changes. After calibrating the transition matrix in the 5-fold opinion space (to calibrate the matrix, the methodology presented in (Kozitsin, 2022) was implemented), one can obtain (subject to two decimal places):

$$P_1 = \begin{bmatrix} .95 & .04 & .01 & 0 & 0 \\ .95 & .04 & .01 & 0 & 0 \\ .94 & .05 & .01 & 0 & 0 \\ .91 & .07 & .02 & 0 & 0 \\ .92 & .06 & .01 & .01 & 0 \end{bmatrix}, P_2 = \begin{bmatrix} .06 & .87 & .07 & 0 & 0 \\ .05 & .9 & .05 & 0 & 0 \\ .03 & .89 & .07 & 0 & 0 \\ .04 & .86 & .1 & 0 & 0 \\ .05 & .86 & .09 & 0 & .005 \end{bmatrix},$$

$$P_3 = \begin{bmatrix} 0 & .06 & .92 & .02 & 0 \\ 0 & .05 & .93 & .01 & 0 \\ 0 & .04 & .94 & .02 & 0 \\ 0 & .04 & .91 & .05 & 0 \\ 0 & .04 & .9 & .06 & 0 \end{bmatrix}, P_4 = \begin{bmatrix} 0 & 0 & .1 & .88 & .02 \\ 0 & 0 & .09 & .89 & .02 \\ 0 & 0 & .08 & .9 & .02 \\ 0 & 0 & .07 & .9 & .03 \\ 0 & 0 & .07 & .88 & .04 \end{bmatrix}, \quad \text{(S3)}$$

$$P_5 = \begin{bmatrix} 0 & 0 & .01 & .1 & .89 \\ 0 & 0 & .01 & .08 & .91 \\ 0 & 0 & .01 & .07 & .92 \\ 0 & 0 & .01 & .07 & .92 \\ 0 & 0 & .01 & .07 & .92 \end{bmatrix}.$$

For the opinion dynamics governed by the transition matrix (S3), it is not a trivial task to predict what control the Person should use to achieve his/her purpose, compared to Scenarios 1 and 2 whereby optimal control strategies are intuitively clear. In this regard, all one can do is rely on the FBS method and, perhaps, compare its output against some trivial controls.

## 5.2. Design of experiments and expectations

To summarize, Scenarios 1 and 2 are characterized by the symmetric transition matrices so one can assume, without loss of generality, that the Person strives to shift agents' opinions towards the left endpoint of the opinion spectrum. Further, it will be assumed that the Person does not care about intermediate states of the social system and his/her objective is only the terminal opinions. Technically, for these two Scenarios, the cost functional with the following opinion-weight vectors is considered:

$$\vec{w} = [0 \quad 0 \quad 0 \quad 0 \quad 0]^T, \vec{v} = [0 \quad 2 \quad 4 \quad 6 \quad 8]^T.$$

In contrast, Scenario 3 is characterized by the asymmetric map of influence processes, so the best control could depend on which side of the opinion spectrum the Person tries to move the system. On this basis, in this case, it is necessary to analyse two inverse configurations of the opinion-weight vectors:

$$\vec{w} = [0 \quad 0 \quad 0 \quad 0 \quad 0]^T, \vec{v} = [0 \quad 2 \quad 4 \quad 6 \quad 8]^T,$$
$$\vec{w} = [0 \quad 0 \quad 0 \quad 0 \quad 0]^T, \vec{v} = [8 \quad 6 \quad 4 \quad 2 \quad 0]^T.$$

The first one means that the Person pushes the system towards the left, while the second refers to pushing toward the right endpoint of the opinion space.

In experiments, the social system starts from three initial opinion distributions: (i) left-peak distribution $(\vec{y}^0 = [n^b \quad 0 \quad 0 \quad 0 \quad 0]^T)$, (ii) uniform distribution $(\vec{y}^0 = \left[\frac{n^b}{5} \quad \frac{n^b}{5} \quad \frac{n^b}{5} \quad \frac{n^b}{5} \quad \frac{n^b}{5}\right]^T)$, and (iii) right-peak distribution $(\vec{y}^0 = [0 \quad 0 \quad 0 \quad 0 \quad n^b]^T)$. By doing so, it is possible to investigate the output of the FBS method for different initial points of the social system including situations when the system is already in a desirable position – control strategies that should be implemented in these cases are referred to as retention strategies. Retention strategies are of particular interest within Scenario 3 whereby dissimilative opinion changes may occur and thus the Person should think not only about how to attract agents' opinions but also about how not to antagonize them. Relying on (González-Bailón & De Domenico, 2021) whereby the authors report that the fraction of bots on Twitter is about 0.4, let us set $n^a = 0.6, n^b = 0.4$. These settings justify the assumption of the huge number of agents of both types in the system. In experiments, let us focus on the time interval $t_0 = 0, t_1 = 40$ and use the grid of 41 points that are defined as:

$$\tau_0 = 0, \tau_1 = 1, \dots, \tau_{39} = 39, \tau_{40} = 40.$$

To present the outputs of the algorithm, the variable $k \in \{0\} \cup [40]$ is used, which denotes concisely the points of the grid.

To test the sensitivity of the algorithm towards the initial guess, let us use five "basic" static controls as starting approximations:

$$\vec{u}^0(\tau) \in \begin{cases} \equiv [n^b \quad 0 \quad 0 \quad 0 \quad 0]^T \\ \equiv [0 \quad n^b \quad 0 \quad 0 \quad 0]^T \\ \equiv [0 \quad 0 \quad n^b \quad 0 \quad 0]^T \\ \equiv [0 \quad 0 \quad 0 \quad n^b \quad 0]^T \\ \equiv [0 \quad 0 \quad 0 \quad 0 \quad n^b]^T \end{cases}$$

In experiments, the algorithm demonstrated two types of limiting behaviour:

1. Convergence – when the FBS method finds a control $\vec{u}^*(\tau)$ at some iteration and then does not leave it through the next iteration.

2. Loop – when the algorithm finds itself in a cycle of $(k)$ approximations:

$$\dots, \underbrace{\vec{u}^{i_1}(\tau), \dots, \vec{u}^{i_k}(\tau)}_{\text{one cycle of length } k}, \vec{u}^{i_1}(\tau), \dots$$

In the examples elaborated below, it usually takes no more than 20 iterations to understand the type of limiting behaviour. Only a few examples need more than 20 iterations to get the output of the algorithm. If the algorithm converges, then its last output ("fixed point") and the best output (the control that is characterized by the lowest value of the objective functional – that does not necessarily coincide with the last output) are displayed. Note that initial guesses are also accounted for when obtaining the best outputs. If the algorithm is cycled, then only the best output is shown.

Further, in this case, the length of the cycle is presented. In all situations, the number of iterations left before stopping is displayed. Besides, special attention is paid to singular controls (see Statement 5) and the situations when this sort of control appears. While presenting results, controls derived by the algorithm with the corresponding values of the objective functional (subject to three decimal places) are shown.

Before turning to examples, let us summarize the expectations regarding the outputs of the FBS method across Scenarios 1–3 (see Table 2).

Table 2.

Expectations from the work of the algorithms across Scenarios 1–3 and different initial opinion distributions

| Time span | Initial opinion distribution $\vec{y}^0$ | | |
|---|---|---|---|
| | Left-peak | Uniform | Right-peak |
| Scenario 1 (pushing towards the left) | $[.4 \quad 0 \quad 0 \quad 0 \quad 0]^T$ | $[.4 \quad 0 \quad 0 \quad 0 \quad 0]^T$ | $[.4 \quad 0 \quad 0 \quad 0 \quad 0]^T$ |
| Scenario 2 (pushing towards the left) | $[.4 \quad 0 \quad 0 \quad 0 \quad 0]^T$ | ? | Some sort of nudging, for example: $[0 \quad 0 \quad 0 \quad 4. \quad 0]^T, \quad k \leq k_1$ $[0 \quad 0 \quad .4 \quad 0 \quad 0]^T, \quad k_1 \leq k \leq k_2$ $[0 \quad .4 \quad 0 \quad 0 \quad 0]^T, \quad k_2 < k \leq k_3$ $[.4 \quad 0 \quad 0 \quad 0 \quad 0]^T, \quad k_3 < k$ |
| Scenario 3 (pushing towards the left) | ? | ? | ? |

| Scenario 3 (pushing towards the left) | ? | ? | ? |
|---|---|---|---|

## 5.3. Results

*Scenario 1*

Numerical experiments (see Tables 3–5, in which the outputs of the algorithm are presented across different initial points of the system and initial guesses) revealed that for the majority of initial guesses, the algorithm correctly identified the optimal control (13). In all situations, the FBS method converges in no more than 5 iterations. Some problems could arise in situations when the algorithm seeks the best retention strategies – see Table 3, initial guesses $[0 \quad 0 \quad .4 \quad 0 \quad 0]^T$ and $[0 \quad 0 \quad 0 \quad .4 \quad 0]^T$. Interestingly, if one applies the initial guess $[0 \quad 0 \quad 0 \quad 0 \quad .4]^T$, which is more distant from the optimal control, then the situation is back to normal. It is worthy to note also that Scenario 1 serves as an example of the non-uniqueness of the optimal control. Within Scenario 1, the system is "well-controlled" in the sense it can be pushed both to the left and right endpoints of the opinion spectrum in finite time regardless of its initial position. As such, for sufficiently long-time intervals, one can expect that small deviations from the optimal control that happens at the beginning of the system evolution should not hamper the situation (the value of the objective functional will be still equal to zero). In general, within Scenario 1, the algorithm is quite robust against the initial guess.

Table 3

Results of numerical experiments within Scenario 1, the left-peak initial opinion distribution $\vec{y}^0 = [n^a \quad 0 \quad 0 \quad 0 \quad 0]^T$

| Initial guess | Last approximation | Best approximation | Convergence details |
|---|---|---|---|
| $[.4 \ 0 \ 0 \ 0 \ 0]^T$ | $[.4 \ 0 \ 0 \ 0 \ 0]^T$ <br><br> 0 | $[.4 \ 0 \ 0 \ 0 \ 0]^T$ <br><br> 0 | Convergence (1 iteration) |
| $[0 \ .4 \ 0 \ 0 \ 0]^T$ | $[.4 \ 0 \ 0 \ 0 \ 0]^T$ <br><br> 0 | $[.4 \ 0 \ 0 \ 0 \ 0]^T$ <br><br> 0 | Convergence (5 iterations) |
| $[0 \ 0 \ .4 \ 0 \ 0]^T$ | $[0 \ 0 \ 0 \ 0 \ .4]^T, \quad k \le 9$ <br> $[.4 \ 0 \ 0 \ 0 \ 0]^T, \quad 9 < k$ <br><br> 1.006 | $[0 \ 0 \ 0 \ 0 \ .4]^T, \quad k \le 7$ <br> $[.4 \ 0 \ 0 \ 0 \ 0]^T, \quad 7 < k$ <br><br> 0.315 | Convergence (4 iterations) |
| $[0 \ 0 \ 0 \ .4 \ 0]^T$ | $[0 \ 0 \ 0 \ 0 \ .4]^T, \quad k \le 9$ <br> $[.4 \ 0 \ 0 \ 0 \ 0]^T, \quad 9 < k$ <br><br> 1.006 | $[0 \ 0 \ 0 \ .4 \ 0]^T, \quad k \le 9$ <br> $[.4 \ 0 \ 0 \ 0 \ 0]^T, \quad 9 < k$ <br><br> 0.425 | Convergence (4 iterations) |
| $[0 \ 0 \ 0 \ 0 \ .4]^T$ | $[.4 \ 0 \ 0 \ 0 \ 0]^T$ <br><br> 0 | $[.4 \ 0 \ 0 \ 0 \ 0]^T$ <br><br> 0 | Convergence (2 iterations) |

Table 4

Results of numerical experiments, the uniform initial opinion distribution $\vec{y}^0 = \left[\frac{n^a}{5} \quad \frac{n^a}{5} \quad \frac{n^a}{5} \quad \frac{n^a}{5} \quad \frac{n^a}{5}\right]^T$

| Initial guess | Last approximation | Best approximation | Convergence details |
|---|---|---|---|
| $[.4 \ 0 \ 0 \ 0 \ 0]^T$ | $[.4 \ 0 \ 0 \ 0 \ 0]^T$ <br><br> 1.224 | $[.4 \ 0 \ 0 \ 0 \ 0]^T$ <br><br> 1.224 | Convergence (1 iteration) |
| $[0 \ .4 \ 0 \ 0 \ 0]^T$ | $[.4 \ 0 \ 0 \ 0 \ 0]^T$ <br><br> 1.224 | $[.4 \ 0 \ 0 \ 0 \ 0]^T$ <br><br> 1.224 | Convergence (5 iterations) |
| $[0 \ 0 \ .4 \ 0 \ 0]^T$ | $[.4 \ 0 \ 0 \ 0 \ 0]^T$ <br><br> 1.224 | $[.4 \ 0 \ 0 \ 0 \ 0]^T$ <br><br> 1.224 | Convergence (2 iterations) |
| $[0 \ 0 \ 0 \ .4 \ 0]^T$ | $[.4 \ 0 \ 0 \ 0 \ 0]^T$ <br><br> 1.224 | $[.4 \ 0 \ 0 \ 0 \ 0]^T$ <br><br> 1.224 | Convergence (2 iterations) |

| | $[.4 \ 0 \ 0 \ 0 \ 0]^T$ | $[.4 \ 0 \ 0 \ 0 \ 0]^T$ | |
| $[0 \ 0 \ 0 \ 0 \ .4]^T$ | | | Convergence (6 iterations) |
| | 1.224 | 1.224 | |

Table 5

Results of numerical experiments, the right-peak initial opinion distribution $\vec{y}^0 = [0 \ 0 \ 0 \ 0 \ n^a]^T$

| Initial guess | Last approximation | Best approximation | Convergence details |
|---|---|---|---|
| $[.4 \ 0 \ 0 \ 0 \ 0]^T$ | $[.4 \ 0 \ 0 \ 0 \ 0]^T$ <br><br> 2.121 | $[.4 \ 0 \ 0 \ 0 \ 0]^T$ <br><br> 2.121 | Convergence (1 iteration) |
| $[0 \ .4 \ 0 \ 0 \ 0]^T$ | $[.4 \ 0 \ 0 \ 0 \ 0]^T$ <br><br> 2.121 | $[.4 \ 0 \ 0 \ 0 \ 0]^T$ <br><br> 2.121 | Convergence (2 iterations) |
| $[0 \ 0 \ .4 \ 0 \ 0]^T$ | $[.4 \ 0 \ 0 \ 0 \ 0]^T$ <br><br> 2.121 | $[.4 \ 0 \ 0 \ 0 \ 0]^T$ <br><br> 2.121 | Convergence (2 iterations) |
| $[0 \ 0 \ 0 \ .4 \ 0]^T$ | $[.4 \ 0 \ 0 \ 0 \ 0]^T$ <br><br> 2.121 | $[.4 \ 0 \ 0 \ 0 \ 0]^T$ <br><br> 2.121 | Convergence (3 iterations) |
| $[0 \ 0 \ 0 \ 0 \ .4]^T$ | $[.4 \ 0 \ 0 \ 0 \ 0]^T$ <br><br> 2.121 | $[.4 \ 0 \ 0 \ 0 \ 0]^T$ <br><br> 2.121 | Convergence (2 iterations) |

*Scenario 2*

Tables 6–8 demonstrate how the algorithm operates under the transition matrix (S2'). If the system is initially located at the left edge of the opinion space, then the algorithm converges in no more than five iterations to the optimal control (see Table 6). For systems that are initially heterogeneous (see Table 7), one can observe that the algorithm either converges or gets caught in a cycle (of the fixed length 15). The optimal strategy found by the FBS method in this case is still control (13). However, the algorithm does not output this result as the last approximation in the case of convergence. Contrary, If the FBS method does not converge, then its best output is

$$[0 \quad 0 \quad .4 \quad 0 \quad 0]^T, \quad k = 0$$
$$[0 \quad .4 \quad 0 \quad 0 \quad 0]^T, \quad 0 < k \le 10$$
$$[.4 \quad 0 \quad 0 \quad 0 \quad 0]^T, \quad 10 < k$$

which is an example of the nudging-type strategy announced before. However, in this particular situation, such control is not optimal (the value of the functional is 1.271 vs 1.238 for control strategy (13)).

Table 6

Results of numerical experiments within Scenario 2, the left-peak initial opinion distribution $\vec{y}^0 = [n^a \quad 0 \quad 0 \quad 0 \quad 0]^T$

| Initial guess | Last approximation | Best approximation | Convergence details |
|---|---|---|---|
| $[.4 \quad 0 \quad 0 \quad 0 \quad 0]^T$ | $[.4 \quad 0 \quad 0 \quad 0 \quad 0]^T$<br><br>0 | $[.4 \quad 0 \quad 0 \quad 0 \quad 0]^T$<br><br>0 | Convergence (1 iteration) |
| $[0 \quad .4 \quad 0 \quad 0 \quad 0]^T$ | $[.4 \quad 0 \quad 0 \quad 0 \quad 0]^T$<br><br>0 | $[.4 \quad 0 \quad 0 \quad 0 \quad 0]^T$<br><br>0 | Convergence (4 iterations) |
| $[0 \quad 0 \quad .4 \quad 0 \quad 0]^T$ | $[.4 \quad 0 \quad 0 \quad 0 \quad 0]^T$<br><br>0 | $[.4 \quad 0 \quad 0 \quad 0 \quad 0]^T$<br><br>0 | Convergence (12 iterations) |
| $[0 \quad 0 \quad 0 \quad .4 \quad 0]^T$ | $[.4 \quad 0 \quad 0 \quad 0 \quad 0]^T$<br><br>0 | $[.4 \quad 0 \quad 0 \quad 0 \quad 0]^T$<br><br>0 | Convergence (12 iterations) |
| $[0 \quad 0 \quad 0 \quad 0 \quad .4]^T$ | $[.4 \quad 0 \quad 0 \quad 0 \quad 0]^T$<br><br>0 | $[.4 \quad 0 \quad 0 \quad 0 \quad 0]^T$<br><br>0 | Convergence (2 iterations) |

Table 7

Results of numerical experiments within Scenario 2, the uniform initial opinion distribution $\vec{y}^0 = \left[\frac{n^a}{5} \quad \frac{n^a}{5} \quad \frac{n^a}{5} \quad \frac{n^a}{5} \quad \frac{n^a}{5}\right]^T$

| Initial guess | Last approximation | Best approximation | Convergence details |
|---|---|---|---|

| Initial guess | Last approximation | Best approximation | Convergence details |
|---|---|---|---|
| $[.4\ 0\ 0\ 0\ 0]^T$ | $[0\ 0\ 0\ .4\ 0]^T,\quad k \le 11$<br>$[.4\ 0\ 0\ 0\ 0]^T,\quad 11 < k$<br><br>1.655 | $[.4\ 0\ 0\ 0\ 0]^T$<br><br>1.238 | Convergence (2 iterations) |
| $[0\ .4\ 0\ 0\ 0]^T$ | | $[0\ 0\ .4\ 0\ 0]^T,\quad k = 0$<br>$[0\ .4\ 0\ 0\ 0]^T,\quad 0 < k \le 10$<br>$[.4\ 0\ 0\ 0\ 0]^T,\quad 10 < k$<br><br>1.271 | Loop (16 iterations, cycle of length 15) |
| $[0\ 0\ .4\ 0\ 0]^T$ | $[0\ 0\ 0\ .4\ 0]^T,\quad k \le 11$<br>$[.4\ 0\ 0\ 0\ 0]^T,\quad 11 < k$<br><br>1.655 | $[.4\ 0\ 0\ 0\ 0]^T$<br><br>1.238 | Convergence (7 iterations) |
| $[0\ 0\ 0\ .4\ 0]^T$ | $[0\ 0\ 0\ .4\ 0]^T,\quad k \le 11$<br>$[.4\ 0\ 0\ 0\ 0]^T,\quad 11 < k$<br><br>1.655 | $[.4\ 0\ 0\ 0\ 0]^T$<br><br>1.238 | Convergence (6 iterations) |
| $[0\ 0\ 0\ 0\ .4]^T$ | | $[0\ 0\ .4\ 0\ 0]^T,\quad k = 0$<br>$[0\ .4\ 0\ 0\ 0]^T,\quad 0 < k \le 10$<br>$[.4\ 0\ 0\ 0\ 0]^T,\quad 10 < k$<br><br>1.271 | Loop (18 iterations, cycle of length 15) |

Table 8

Results of numerical experiments within Scenario 2, the right-peak initial opinion distribution $\vec{y}^0 = [0\ \ 0\ \ 0\ \ 0\ \ n^a]^T$

| Initial guess | Last approximation | Best approximation | Convergence details |
|---|---|---|---|
| $[.4\ 0\ 0\ 0\ 0]^T$ | | $[0\ 0\ .4\ 0\ 0]^T,\quad k \le 19$<br>$[0\ .4\ 0\ 0\ 0]^T,\quad 19 < k \le 37$<br>$[.4\ 0\ 0\ 0\ 0]^T,\quad 37 < k$<br><br>3.129 | Loop (117 iterations, cycle of length 88) |
| $[0\ .4\ 0\ 0\ 0]^T$ | | $[0\ 0\ .4\ 0\ 0]^T,\quad k \le 19$<br>$[0\ .4\ 0\ 0\ 0]^T,\quad 19 < k \le 37$<br>$[.4\ 0\ 0\ 0\ 0]^T,\quad 37 < k$<br><br>3.129 | Loop (95 iterations, cycle of length 88) |
| $[0\ 0\ .4\ 0\ 0]^T$ | | $[0\ 0\ .4\ 0\ 0]^T,\quad k \le 19$<br>$[0\ .4\ 0\ 0\ 0]^T,\quad 19 < k \le 37$<br>$[.4\ 0\ 0\ 0\ 0]^T,\quad 37 < k$ | Loop (123 iterations, cycle of length 88) |
| | | 3.129 | |
| $[0 \quad 0 \quad 0 \quad .4 \quad 0]^T$ | | $\begin{aligned}[0 \quad 0 \quad .4 \quad 0 \quad 0]^T, &\quad k \leq 19\\ [0 \quad .4 \quad 0 \quad 0 \quad 0]^T, &\quad 19 < k \leq 37\\ [.4 \quad 0 \quad 0 \quad 0 \quad 0]^T, &\quad 37 < k\end{aligned}$ 3.129 | Loop (101 iterations, cycle of length 88) |
| $[0 \quad 0 \quad 0 \quad 0 \quad .4]^T$ | | $\begin{aligned}[0 \quad 0 \quad .4 \quad 0 \quad 0]^T, &\quad k \leq 19\\ [0 \quad .4 \quad 0 \quad 0 \quad 0]^T, &\quad 19 < k \leq 37\\ [.4 \quad 0 \quad 0 \quad 0 \quad 0]^T, &\quad 37 < k\end{aligned}$ 3.129 | Loop (118 iterations, cycle of length 88) |

If one considers the system initially located on the right endpoint of the opinion spectrum (see Table 8), then the situation changes. The convergence ability of the algorithm is seriously hampered: it does not converge and needs ~100 or more iterations to reveal the type of limiting behaviour. However, the best strategy derived from the FBS method, in this case, is

$$[0 \quad 0 \quad .4 \quad 0 \quad 0]^T, \quad k \leq 19$$
$$[0 \quad .4 \quad 0 \quad 0 \quad 0]^T, \quad 19 < k \leq 37$$
$$[.4 \quad 0 \quad 0 \quad 0 \quad 0]^T, \quad 37 < k$$

and does not depend on the initial guess and perfectly coincides with the expectations (see Table 2). One more interesting detail is that for the initial guess $\vec{u}^0 = [0 \quad 0 \quad 0 \quad 0 \quad .4]^T$ the algorithm comes to a singular control in which derivatives $\frac{\partial H(\tau_k)}{\partial u_2}$ and $\frac{\partial H(\tau_k)}{\partial u_3}$ have the highest value at the point $\tau_k = 40$ of the grid. In this case, system (12) suggests the following solution:

$$\vec{u}(\tau_k) = [0 \quad 0 \quad .4 \quad 0 \quad 0]^T.$$

For Scenario 2, one can notice that depending on the initial position of the system, the algorithm may be robust (left- and right-peak opinion distributions) or not (the uniform initial opinion distribution) against the initial guess.

*Scenario 3*

As was mentioned before, in the case of Scenario 3, one cannot check whether the output of the algorithm is optimal and, if not, how it differs from the optimal one. In this regard, all one can do is interpret the results. However, Scenario 3 has one striking difference, compared to Scenarios 1 and 2: it is drawn upon the asymmetric transition matrix (S3). As such, one can compare strategies, obtained by the FBS method in case the Person strives to push agents' opinions towards the left against those derived for the situation when the Person wants to move the system towards the right endpoint of the opinion spectrum. Within Scenario 3, the algorithm typically needs only a few iterations to give an output. Further, this Scenario is characterized by relatively weak robustness of the algorithm against the initial guesses – sometimes it is essential to consider a few initial guesses to obtain an acceptable output.

Let us start with a question about how the Person should act to push the system towards the left (see Tables 9–11). To summarize, if it comes to dealing with initially heterogeneous systems or systems that are already in the desired position $\vec{y}^0 = [n^a \quad 0 \quad 0 \quad 0 \quad 0]^T$ (see Tables 9, 10), then the best strategy obtained by the algorithm is to translate opinion $x_2$. That is, instead of using the "strict" control (13), the Person should apply the softer approach. To understand this feature, let us take a closer look at the transition matrix (S3). More precisely, it is necessary to analyse its first slice over the first axis (for now, it is presented as subject to three decimal places):

$$P_{1,:,:} := \begin{bmatrix} .947 & .044 & .008 & .001 & 0 \\ .954 & .04 & .005 & 0 & 0 \\ .938 & .051 & .01 & .001 & 0 \\ .911 & .07 & .016 & .003 & 0 \\ .919 & .063 & .009 & .009 & 0 \end{bmatrix}. \tag{S4}$$

Table 9

Results of numerical experiments within Scenario 3, the left-peak initial opinion distribution $\vec{y}^0 = [n^a \quad 0 \quad 0 \quad 0 \quad 0]^T$, $\vec{w} = [0 \quad 0 \quad 0 \quad 0 \quad 0]^T$, $\vec{v} = [0 \quad 2 \quad 4 \quad 6 \quad 8]^T$ (pushing the system towards the left)

| Initial guess | Last approximation | Best approximation | Convergence details |
|---|---|---|---|
| $[.4 \quad 0 \quad 0 \quad 0 \quad 0]^T$ | $[0 \quad 0 \quad 0 \quad 0 \quad .4]^T, \quad k \leq 8$ <br> $[0 \quad .4 \quad 0 \quad 0 \quad 0]^T, \quad 8 < k$ | $[.4 \quad 0 \quad 0 \quad 0 \quad 0]^T$ <br><br> $1.404$ | Convergence (2 iterations) |

| Initial guess | Last approximation | Best approximation | Convergence details |
|---|---|---|---|
| | 1.422 | | |
| $[0 \ \ .4 \ \ 0 \ \ 0 \ \ 0]^T$ | | $[0 \ \ .4 \ \ 0 \ \ 0 \ \ 0]^T$ <br><br> 1.31 | Loop (6 iterations, cycle of length 5) |
| $[0 \ \ 0 \ \ .4 \ \ 0 \ \ 0]^T$ | $[0 \ \ 0 \ \ 0 \ \ 0 \ \ .4]^T, \quad k \leq 8$ <br> $[0 \ \ .4 \ \ 0 \ \ 0 \ \ 0]^T, \quad 8 < k$ <br><br> 1.422 | $[0 \ \ 0 \ \ 0 \ \ 0 \ \ .4]^T, \quad k \leq 8$ <br> $[0 \ \ .4 \ \ 0 \ \ 0 \ \ 0]^T, \quad 8 < k$ <br><br> 1.422 | Convergence (2 iterations) |
| $[0 \ \ 0 \ \ 0 \ \ .4 \ \ 0]^T$ | | $[0 \ \ .4 \ \ 0 \ \ 0 \ \ 0]^T$ <br><br> 1.31 | Loop (9 iterations, cycle of length 5) |
| $[0 \ \ 0 \ \ 0 \ \ 0 \ \ .4]^T$ | | $[0 \ \ .4 \ \ 0 \ \ 0 \ \ 0]^T$ <br><br> 1.31 | Loop (13 iterations, cycle of length 5) |

Table 10

Results of numerical experiments within Scenario 3, the uniform initial opinion distribution $\vec{y}^0 = \left[\frac{n^a}{5} \ \ \frac{n^a}{5} \ \ \frac{n^a}{5} \ \ \frac{n^a}{5} \ \ \frac{n^a}{5}\right]^T$, $\vec{w} = [0 \ \ 0 \ \ 0 \ \ 0 \ \ 0]^T$, $\vec{v} = [0 \ \ 2 \ \ 4 \ \ 6 \ \ 8]^T$ (pushing the system towards the left)

| Initial guess | Last approximation | Best approximation | Convergence details |
|---|---|---|---|
| $[.4 \ \ 0 \ \ 0 \ \ 0 \ \ 0]^T$ | | $[0 \ \ .4 \ \ 0 \ \ 0 \ \ 0]^T$ <br><br> 1.797 | Loop (15 iterations, cycle of length 7) |
| $[0 \ \ .4 \ \ 0 \ \ 0 \ \ 0]^T$ | | $[0 \ \ .4 \ \ 0 \ \ 0 \ \ 0]^T$ <br><br> 1.797 | Loop (8 iterations, cycle of length 7) |
| $[0 \ \ 0 \ \ .4 \ \ 0 \ \ 0]^T$ | $[0 \ \ 0 \ \ 0 \ \ 0 \ \ .4]^T, \quad k \leq 11$ <br> $[0 \ \ .4 \ \ 0 \ \ 0 \ \ 0]^T, \quad 11 < k$ <br><br> 1.879 | $[0 \ \ 0 \ \ 0 \ \ .4 \ \ 0]^T, \quad k \leq 7$ <br> $[0 \ \ 0 \ \ .4 \ \ 0 \ \ 0]^T, \quad 7 < k \leq 11$ <br> $[0 \ \ .4 \ \ 0 \ \ 0 \ \ 0]^T, \quad 11 < k$ <br><br> 1.859 | Convergence (4 iterations) |
| $[0 \ \ 0 \ \ 0 \ \ .4 \ \ 0]^T$ | | $[0 \ \ .4 \ \ 0 \ \ 0 \ \ 0]^T$ <br><br> 1.797 | Loop (11 iterations, cycle of length 7) |
| $[0 \ \ 0 \ \ 0 \ \ 0 \ \ .4]^T$ | | $[0 \ \ .4 \ \ 0 \ \ 0 \ \ 0]^T$ | Loop (9 iterations, cycle of length 7) |

| | | | 1.797 | |
| --- | --- | --- | --- | --- |

Table 11

Results of numerical experiments within Scenario 3, the right-peak initial opinion distribution $\vec{y}^0 = [0 \quad 0 \quad 0 \quad 0 \quad n^a]^T$, $\vec{w} = [0 \quad 0 \quad 0 \quad 0 \quad 0]^T$, $\vec{v} = [0 \quad 2 \quad 4 \quad 6 \quad 8]^T$ (pushing the system towards the left)

| Initial guess | Last approximation | Best approximation | | Convergence details |
| --- | --- | --- | --- | --- |
| $[.4 \quad 0 \quad 0 \quad 0 \quad 0]^T$ | | $[.4 \quad 0 \quad 0 \quad 0 \quad 0]^T,$  $[0 \quad .4 \quad 0 \quad 0 \quad 0]^T,$  2.361 | $k \leq 18$  $18 < k$ | Loop (12 iterations, cycle of length 2) |
| $[0 \quad .4 \quad 0 \quad 0 \quad 0]^T$ | | $[.4 \quad 0 \quad 0 \quad 0 \quad 0]^T,$  $[0 \quad .4 \quad 0 \quad 0 \quad 0]^T,$  2.361 | $k \leq 18$  $18 < k$ | Loop (11 iterations, cycle of length 2) |
| $[0 \quad 0 \quad .4 \quad 0 \quad 0]^T$ | | $[.4 \quad 0 \quad 0 \quad 0 \quad 0]^T,$  $[0 \quad .4 \quad 0 \quad 0 \quad 0]^T,$  2.361 | $k \leq 18$  $18 < k$ | Loop (14 iterations, cycle of length 2) |
| $[0 \quad 0 \quad 0 \quad .4 \quad 0]^T$ | | $[.4 \quad 0 \quad 0 \quad 0 \quad 0]^T,$  $[0 \quad .4 \quad 0 \quad 0 \quad 0]^T,$  2.362 | $k \leq 19$  $19 < k$ | Loop (7 iterations, cycle of length 2) |
| $[0 \quad 0 \quad 0 \quad 0 \quad .4]^T$ | | $[.4 \quad 0 \quad 0 \quad 0 \quad 0]^T,$  $[0 \quad .4 \quad 0 \quad 0 \quad 0]^T,$  2.362 | $k \leq 19$  $19 < k$ | Loop (13 iterations, cycle of length 2) |

From slice (S4), one can notice that agents espousing opinion $x_1$ are more likely to change their opinions if exposed to opinion $x_2$ than if being influenced by $x_1$ (the probability of opinion change 0.954 vs 0.947). This issue explains the output of the algorithm.

If the Person needs to convince individuals espousing opinion $x_5$ which is completely opposite to the desirable one $x_1$ (see Table 11), then he/she has to employ somewhat opposite to nudging: use the strict control $[.4 \quad 0 \quad 0 \quad 0 \quad 0]^T$ at first and then, when the system becomes closer to the left endpoint of the opinion spectrum, switch the control to the opinion $[0 \quad .4 \quad 0 \quad 0 \quad 0]^T$ which suits better while dealing individuals espousing opinions near to $x_1$.

What is interesting, if the Person aims to push agents' opinions towards the right edge of the opinion spectrum (see Tables 12–14), then the situation becomes completely different. If the purpose of the Person is to convince individuals espousing the opposite opinion ($x_1$ – see Table 12), then a sort of nudging is the best answer of the FBS method:

$$[0 \quad 0 \quad 0 \quad .4 \quad 0]^T, \qquad k \leq 21$$
$$[0 \quad 0 \quad 0 \quad 0 \quad .4]^T, \qquad 21 < k$$

Table 12

Results of numerical experiments within Scenario 3, the left-peak initial opinion distribution $\vec{y}^0 = [n^a \quad 0 \quad 0 \quad 0 \quad 0]^T$, $\vec{w} = [0 \quad 0 \quad 0 \quad 0 \quad 0]^T$, $\vec{v} = [8 \quad 6 \quad 4 \quad 2 \quad 0]^T$ (pushing the system towards the right)

| Initial guess | Last approximation | Best approximation | Convergence details |
|---|---|---|---|
| $[.4 \quad 0 \quad 0 \quad 0 \quad 0]^T$ | | $\begin{array}{ll}[0 \quad .4 \quad 0 \quad 0 \quad 0]^T, & k \leq 5 \\ [.4 \quad 0 \quad 0 \quad 0 \quad 0]^T, & 5 < k \leq 9 \\ [0 \quad 0 \quad .4 \quad 0 \quad 0]^T, & k = 10 \\ [0 \quad 0 \quad 0 \quad .4 \quad 0]^T, & 10 < k \leq 33 \\ [0 \quad 0 \quad 0 \quad 0 \quad .4]^T, & 33 < k \end{array}$ <br><br> 3.104 | Loop (7 iterations, cycle of length 2) |
| $[0 \quad .4 \quad 0 \quad 0 \quad 0]^T$ | | $\begin{array}{ll}[0 \quad 0 \quad 0 \quad .4 \quad 0]^T, & k \leq 21 \\ [0 \quad 0 \quad 0 \quad 0 \quad .4]^T, & 21 < k \end{array}$ <br><br> 2.977 | Loop (10 iterations, cycle of length 2) |
| $[0 \quad 0 \quad .4 \quad 0 \quad 0]^T$ | | $\begin{array}{ll}[0 \quad 0 \quad 0 \quad .4 \quad 0]^T, & k \leq 7 \\ [0 \quad 0 \quad 0 \quad 0 \quad .4]^T, & 7 < k \leq 17 \\ [0 \quad 0 \quad 0 \quad .4 \quad 0]^T, & 17 < k \leq 21 \\ [0 \quad 0 \quad 0 \quad 0 \quad .4]^T, & 21 < k \end{array}$ <br><br> 2.978 | Loop (22 iterations, cycle of length 2) |
| $[0 \quad 0 \quad 0 \quad .4 \quad 0]^T$ | | $\begin{array}{ll}[0 \quad 0 \quad 0 \quad .4 \quad 0]^T, & k \leq 29 \\ [0 \quad 0 \quad 0 \quad 0 \quad .4]^T, & 29 < k \end{array}$ <br><br> 2.978 | Loop (20 iterations, cycle of length 2) |

| Initial guess | Last approximation | Best approximation | Convergence details |
|---|---|---|---|
| $[0 \quad 0 \quad 0 \quad 0 \quad .4]^T$ | | $\begin{array}{ll}[0 \ 0 \ 0 \ .4 \ 0]^T, & k \le 27 \\ {[0 \ 0 \ 0 \ 0 \ .4]^T,} & 27 < k\end{array}$ <br><br> 2.978 | Loop (38 iterations, cycle of length 2) |

Table 13

Results of numerical experiments within Scenario 3, the uniform initial opinion distribution $\vec{y}^0 = \left[\frac{n^a}{5} \quad \frac{n^a}{5} \quad \frac{n^a}{5} \quad \frac{n^a}{5} \quad \frac{n^a}{5}\right]^T$, $\vec{w} = [0 \quad 0 \quad 0 \quad 0 \quad 0]^T$, $\vec{v} = [8 \quad 6 \quad 4 \quad 2 \quad 0]^T$ (pushing the system towards the right)

| Initial guess | Last approximation | Best approximation | Convergence details |
|---|---|---|---|
| $[.4 \quad 0 \quad 0 \quad 0 \quad 0]^T$ | | $\begin{array}{ll}[0 \ 0 \ 0 \ .4 \ 0]^T, & k \le 8 \\ {[0 \ 0 \ .4 \ 0 \ 0]^T,} & 8 < k \le 10 \\ {[0 \ 0 \ 0 \ .4 \ 0]^T,} & 10 < k < 12 \\ {[0 \ 0 \ 0 \ 0 \ .4]^T,} & 12 < k\end{array}$ <br><br> 2.493 | Loop (7 iterations, cycle of length 2) |
| $[0 \quad .4 \quad 0 \quad 0 \quad 0]^T$ <br><br> $[0 \quad 0 \quad 0 \quad 0 \quad .4]^T$ <br><br> 2.481 | | $\begin{array}{ll}[0 \ 0 \ 0 \ 0 \ .4]^T, & k \le 7 \\ {[0 \ 0 \ 0 \ .4 \ 0]^T,} & 7 < k \le 9 \\ {[0 \ 0 \ 0 \ 0 \ .4]^T,} & 9 < k\end{array}$ <br> or <br> $[0 \quad 0 \quad 0 \quad 0 \quad .4]^T$ <br><br> 2.481 | Convergence (7 iterations) |
| $[0 \quad 0 \quad .4 \quad 0 \quad 0]^T$ | | $\begin{array}{ll}[.4 \ 0 \ 0 \ 0 \ 0]^T, & k \le 2 \\ {[0 \ .4 \ 0 \ 0 \ 0]^T,} & 2 < k \le 8 \\ {[0 \ 0 \ .4 \ 0 \ 0]^T,} & 8 < k \le 11 \\ {[0 \ 0 \ 0 \ .4 \ 0]^T,} & 11 < k \le 13 \\ {[0 \ 0 \ 0 \ 0 \ .4]^T,} & 13 < k\end{array}$ <br><br> 2.55 | Loop (4 iterations, cycle of length 2) |
| $[0 \quad 0 \quad 0 \quad .4 \quad 0]^T$ | $[0 \quad 0 \quad 0 \quad 0 \quad .4]^T$ <br><br> 2.481 | $[0 \quad 0 \quad 0 \quad 0 \quad .4]^T$ <br><br> 2.481 | Convergence (3 iterations) |
| $[0 \quad 0 \quad 0 \quad 0 \quad .4]^T$ | $[0 \quad 0 \quad 0 \quad 0 \quad .4]^T$ <br><br> 2.481 | $[0 \quad 0 \quad 0 \quad 0 \quad .4]^T$ <br><br> 2.481 | Convergence (1 iteration) |

Table 14

Results of numerical experiments within Scenario 3, the right-peak initial opinion distribution $\vec{y}^0 = [0 \quad 0 \quad 0 \quad 0 \quad n^a]^T$, $\vec{w} = [0 \quad 0 \quad 0 \quad 0 \quad 0]^T$, $\vec{v} = [8 \quad 6 \quad 4 \quad 2 \quad 0]^T$ (pushing the system towards the right)

| Initial guess | Last approximation | Best approximation | Convergence details |
|---|---|---|---|
| $[.4 \quad 0 \quad 0 \quad 0 \quad 0]^T$ | $[0 \quad 0 \quad 0 \quad .4 \quad 0]^T, \quad k = 0$<br>$[0 \quad 0 \quad 0 \quad 0 \quad .4]^T, \quad 0 < k$<br><br>1.958 | $[0 \quad 0 \quad 0 \quad .4 \quad 0]^T, \quad k = 0$<br>$[0 \quad 0 \quad 0 \quad 0 \quad .4]^T, \quad 0 < k$<br><br>1.958 | Convergence (2 iterations) |
| $[0 \quad .4 \quad 0 \quad 0 \quad 0]^T$ | $[0 \quad 0 \quad 0 \quad .4 \quad 0]^T, \quad k = 0$<br>$[0 \quad 0 \quad 0 \quad 0 \quad .4]^T, \quad 0 < k$<br><br>1.958 | $[0 \quad 0 \quad 0 \quad .4 \quad 0]^T, \quad k = 0$<br>$[0 \quad 0 \quad 0 \quad 0 \quad .4]^T, \quad 0 < k$<br><br>1.958 | Convergence (2 iterations) |
| $[0 \quad 0 \quad .4 \quad 0 \quad 0]^T$ | $[.4 \quad 0 \quad 0 \quad 0 \quad 0]^T, \quad k = 0$<br>$[0 \quad 0 \quad 0 \quad 0 \quad .4]^T, \quad 0 < k$<br><br>1.967 | $[.4 \quad 0 \quad 0 \quad 0 \quad 0]^T, \quad k = 0$<br>$[0 \quad 0 \quad 0 \quad 0 \quad .4]^T, \quad 0 < k$<br><br>1.967 | Convergence (2 iterations) |
| $[0 \quad 0 \quad 0 \quad .4 \quad 0]^T$ | $[.4 \quad 0 \quad 0 \quad 0 \quad 0]^T, \quad k = 0$<br>$[0 \quad 0 \quad 0 \quad 0 \quad .4]^T, \quad 0 < k$<br><br>1.967 | $[.4 \quad 0 \quad 0 \quad 0 \quad 0]^T, \quad k = 0$<br>$[0 \quad 0 \quad 0 \quad 0 \quad .4]^T, \quad 0 < k$<br><br>1.967 | Convergence (2 iterations) |
| $[0 \quad 0 \quad 0 \quad 0 \quad .4]^T$ | $[0 \quad 0 \quad 0 \quad .4 \quad 0]^T, \quad k = 0$<br>$[0 \quad 0 \quad 0 \quad 0 \quad .4]^T, \quad 0 < k$<br><br>1.958 | $[0 \quad 0 \quad 0 \quad 0 \quad .4]^T$<br><br>1.958 | Convergence (2 iterations) |

In turn, Tables 13 and 14 indicate that one should translate the "strict" opinion $x_5$ to convince individuals whose views are sufficiently close to the right edge of the opinion spectrum.

### 5.4. Summary of numerical experiments

Table 15 summarizes the best controls obtained by the algorithm across Scenarios and different initial points of the system, subject to multi starts. One can conclude that the outputs of the FBS method coincide with the expectations, whenever it is possible to compare. However, in some situations, without multi starts, one risks missing the best control strategy. Further, not all the runs of the algorithm end up in fixed points – sometimes, the algorithm is locked in cycles. The absence of convergence may be referred to as the inaccuracy of the numerical approach. Besides, varying the step of the grid does not change the type of limiting behaviour. Observing the outputs of the algorithm, one can conclude that it suggests the use of bang-bang control functions. In fact, this result is a direct consequence of the linearity of the system with respect to the control variables and the Pontryagin maximum principle. Further, no singular control situations in the algorithm outputs were found.

Table 15

The best outputs of the algorithm across Scenarios 1–3 and different values of $\vec{y}^0$

| Time span | Initial opinion distribution $\vec{y}^0$ | | |
|---|---|---|---|
| | Left-peak | Uniform | Right-peak |
| Scenario 1 (pushing towards the left) | $[.4 \quad 0 \quad 0 \quad 0 \quad 0]^T$<br><br>Convergence<br><br>+ | $[.4 \quad 0 \quad 0 \quad 0 \quad 0]^T$<br><br>Convergence<br><br>+ | $[.4 \quad 0 \quad 0 \quad 0 \quad 0]^T$<br><br>Convergence<br><br>+ |
| Scenario 2 (pushing towards the left) | $[.4 \quad 0 \quad 0 \quad 0 \quad 0]^T$<br><br>Convergence<br><br>+ | $[.4 \quad 0 \quad 0 \quad 0 \quad 0]^T$<br><br>Convergence | $[0 \quad 0 \quad .4 \quad 0 \quad 0]^T, \quad k \leq 19$<br>$[0 \quad .4 \quad 0 \quad 0 \quad 0]^T, \quad 19 < k \leq 37$<br>$[.4 \quad 0 \quad 0 \quad 0 \quad 0]^T, \quad 37 < k$<br><br>Loop<br><br>+ |
| Scenario 3 (pushing towards the left) | $[0 \quad .4 \quad 0 \quad 0 \quad 0]^T$<br><br>Loop | $[0 \quad .4 \quad 0 \quad 0 \quad 0]^T$<br><br>Loop | $[.4 \quad 0 \quad 0 \quad 0 \quad 0]^T, \quad k \leq 18$<br>$[0 \quad .4 \quad 0 \quad 0 \quad 0]^T, \quad 18 < k$<br><br>Loop |

| Scenario 3 (pushing towards the right) | $[0 \ \ 0 \ \ 0 \ \ .4 \ \ 0]^T, \quad k \le 21$<br>$[0 \ \ 0 \ \ 0 \ \ 0 \ \ .4]^T, \quad 21 < k$<br><br>Loop | $[0 \ \ 0 \ \ 0 \ \ 0 \ \ .4]^T$<br><br>Convergence | $[0 \ \ 0 \ \ 0 \ \ 0 \ \ .4]^T$<br><br>Convergence |
|---|---|---|---|

Note: each control is accompanied by the corresponding type of limiting behaviour (convergence or loop). Whenever it is possible, pluses or minuses signify whether the algorithm outputs meet the expectations.

## 6. Optimal control in structured populations

The results presented in Sections 4 and 5 were obtained within the mean-field approximation framework that assumes a fully mixed population of agents. However, real social systems deviate from this assumption and are characterized by more complex communication structures. Social graphs are typically sparse, clustered (are organized into cohesive communities), have a relatively short average path length (which is on the order of $\ln N$, where $N$ is the size of the network – six degrees of separation theory), and demonstrate a power-law degree distribution (Newman, 2010). In this regard, one may hypothesize that the results obtained in Sections 4 and 5 are no longer valid for structured populations. However, that is not exactly the case, and the proposed theoretical predictions may be applied to some complex networks, such as Barabasi-Albert graphs, Watts-Strogatz graphs, and random geometric graphs, for some transition matrices. Figure 1 demonstrates a simulation run for the social system of $N = 2000$ agents connected with a random geometric network and influenced by control derived from the algorithm within Scenario 3. As can be seen from Figure 1, the mean-field approximation perfectly predicts the real evolution of the system.

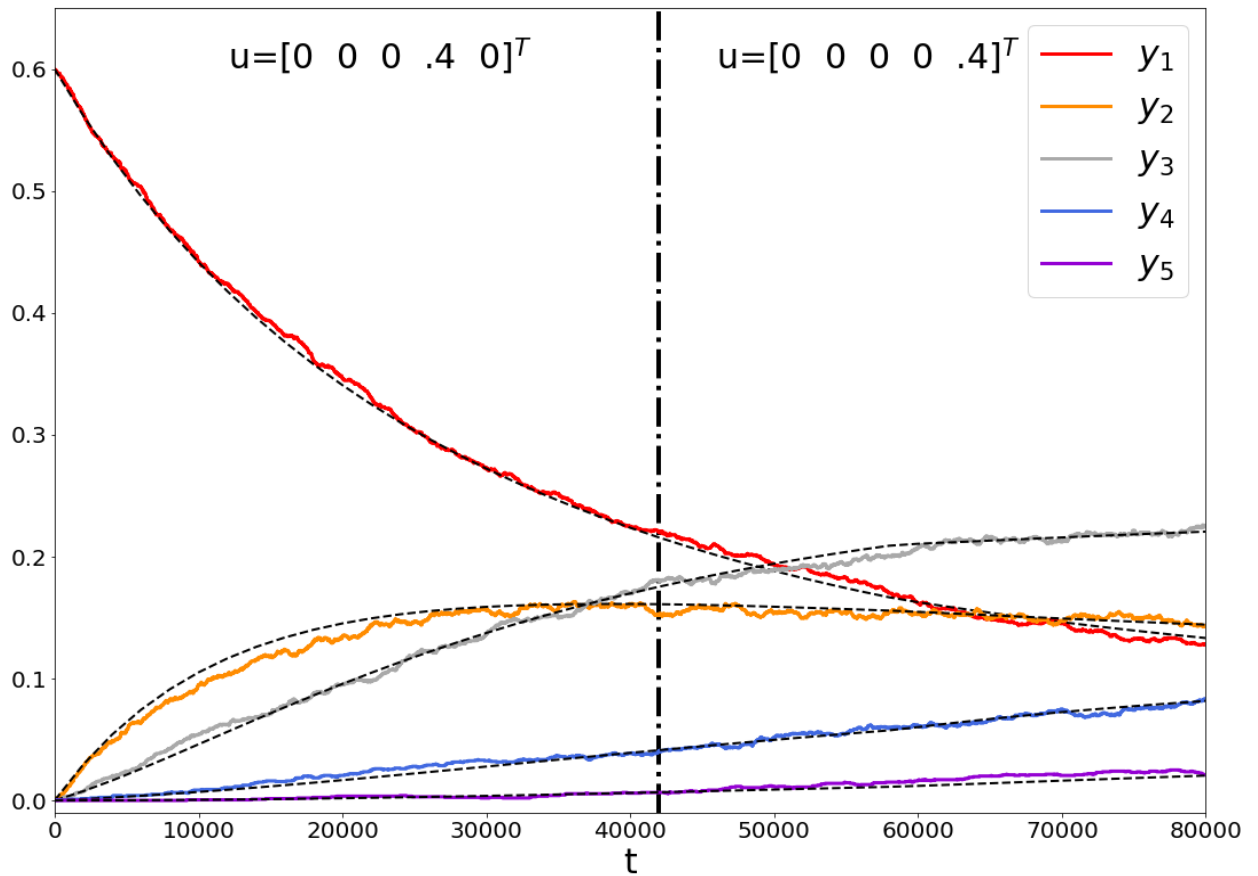

Figure 1. A simulation run of a social system of $N = 2000$ agents connected with a random geometric network (with the threshold value of 0.025). $n^b = 0.4$ of agents are stubborn ones. They obey the following control strategy

$$[0 \quad 0 \quad 0 \quad .4 \quad 0]^T, \qquad k \leq 21$$
$$[0 \quad 0 \quad 0 \quad 0 \quad .4]^T, \qquad 21 < k$$

which was derived by the algorithm (see Scenario 3, the cost functional $\vec{w} = [0 \quad 0 \quad 0 \quad 0 \quad 0]^T, \vec{v} = [8 \quad 6 \quad 4 \quad 2 \quad 0]^T$, and the left-peak initial opinion distribution). Stubborn agents are located in the network purely at random. One step of the grid corresponds to the halfway point of the scaled time $\tau$ and to 2000 points of the initial time $t$ (scaled time is defined as $\tau = \frac{t}{N}$). Coloured lines plot the evolution of the system obtained via the simulation run, while dashed lines depict the mean-field predictions. The vertical dashed line stands for the time moment when the control strategy is switched ($k = 21$).

However, the agreement between theoretical predictions and simulations grows weak if considering, for example, transition matrices from Scenarios 1 or 2 (see Figure 2). The reason is that the underlying transition matrices (S1) and (S2') include some rows that encode deterministic probability distributions (when the output of opinion interaction is unambiguously determined by what opinions interact). As was reported in (Kozitsin, 2022), for such situations the mean-field predictions tend to diverge with the system behaviour.

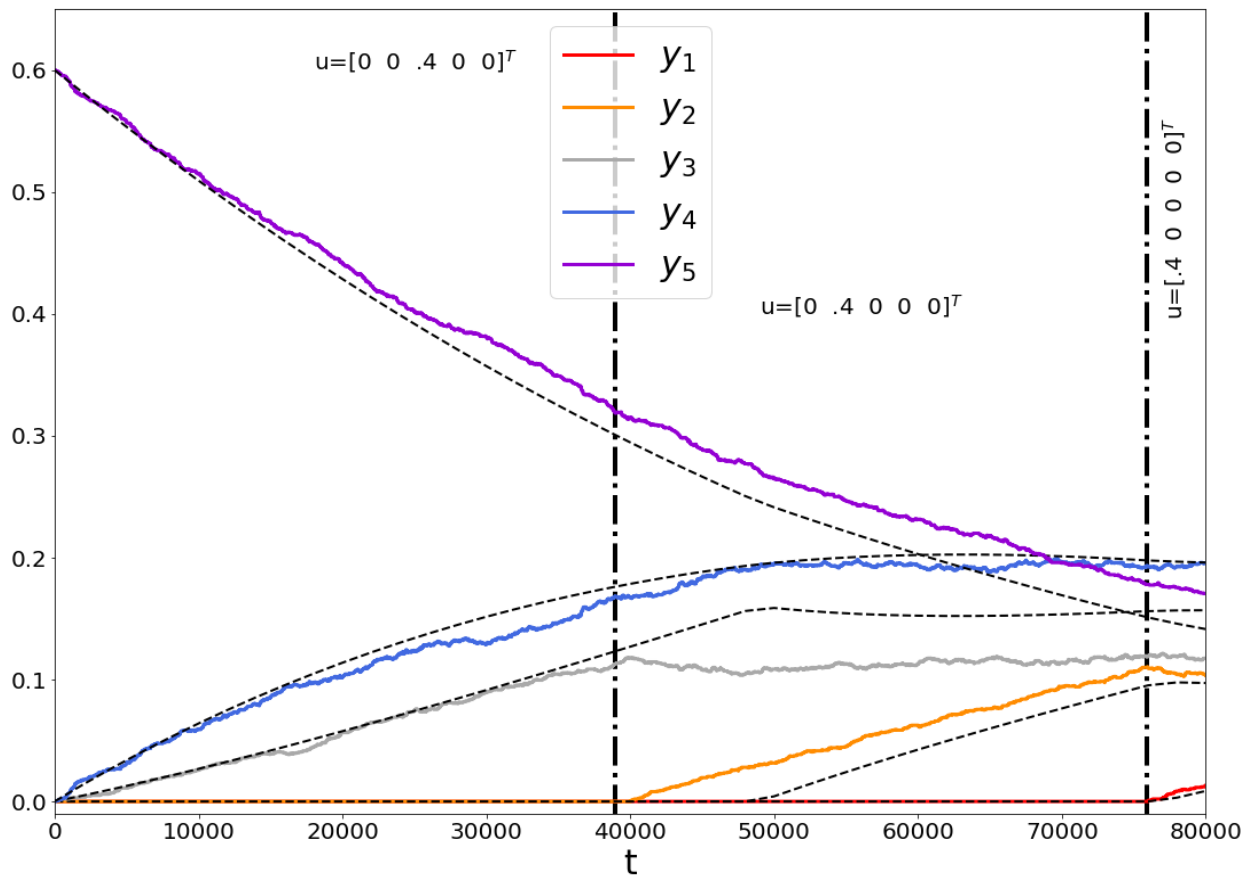

Figure 2. A simulation run of a social system of $N = 2000$ agents connected with a random geometric network (with the threshold value of 0.025). $n^b = 0.4$ of agents are stubborn ones. They obey the control strategy that was obtained by the algorithm within Scenario 2 (the right-peak initial opinion distribution). Coloured lines plot the system evolution obtained via the simulation run, dashed lines depict the mean-field predictions. One step of the grid corresponds to the halfway point of the scaled time $\tau$ and to 2000 points of the initial time $t$ (scaled time is defined as $\tau = \frac{t}{N}$). The vertical dashed line stands for the time moment when the control strategy is switched.

## 7. Conclusion

This article advances the theory of opinion dynamics models by elaborating a quite flexible framework to discern the optimal control in opinion dynamics models. First, the paper proposes a modification of the recently published opinion dynamics model (Kozitsin, 2022). The advanced model accounts for stubborn agents (bots, mass media) that may influence ordinary agents (users). For the advanced model, the mean-field approximation in the form of the system of ordinary differential equations was derived. Some basic properties of this system were investigated. Next, building upon this mean-field approximation, the optimal control problem was formulated and some of its possible generalizations were discussed. For the optimal control problem, the Pontryagin maximum principle was applied, and some properties of adjoint functions and optimal control functions were characterized. After that, using the forward-backward sweep method, it was demonstrated how the elaborated framework could cover the principal microscopic assumptions regarding social influence. It was shown that the algorithm derived insightful control strategies that were correct whenever it was possible to check them. For example, under the assimilative influence assumption (interactions between agents reduce the absolute difference in opinions and the greater initial opinion discrepancies lead to more distant opinion shifts), the algorithm suggests using the most radical left-side opinion, if the purpose is to shift the system towards the left side of the opinion spectrum. Further, under the bounded confidence assumption (assimilative influence may occur only if the opinions of interacting agents are not too distant – otherwise, the recipient rejects the influence and remains of the same opinion), the algorithm successfully reveals that some sort of nudging is required to affect the individuals espousing an opinion contrary to those who should be convinced. Further, it was shown that the framework elaborated in this paper could be applied in some cases to structured populations connected by complex social graphs.

The current situation in the field of opinion dynamics models is that one still does not know exactly how social influence changes individuals' opinions and behaviour. The question of which model best describes real social processes remains debatable. Perhaps, there is no unique answer; in some situations, one employs one approach, while other situations require a different one (Proskurnikov & Tempo, 2018). In this regard, the framework proposed in the current paper is quite robust against this uncertainty because it can be adapted and applied to a huge scope of social influence mechanisms and thus, within this framework, one can approximate a wide range of different opinion formation models.

The promising directions for future studies include testing more different transition matrices derived both from the theoretical models and from empirics, as well as considering two or more concerned persons who want to control the opinions of ordinary individuals – it could be, for example, concurrent firms that fight for customers. In this situation, one should apply the approaches from the theory of differential games. Further, it would be interesting to find optimal control in settings whereby personalization systems are presented (De Marzo *et al*., 2020; Peralta *et al*., 2021). Besides, one can also account for the dynamics of the social graph (Frasca *et al*., 2019; Holme & Newman, 2006) – however, in this case, the mean-field approach is not applicable. Instead, the pair approximation should be employed to get a more precise estimation of the local properties of the system (Jędrzejewski, 2017).

**8. Data availability**

Numerical experiments were conducted in JupiterHub using the Python 3 language. All of the data, codes, and other support information can be found at https://dataverse.harvard.edu/privateurl.xhtml?token=4a07a8b8-2e84-4e30-a2f1-4636d9ce6e3e (Online Supplementary Materials). This is a private URL. If the paper will be accepted for publication, then the online dataset will be published, and this private URL will be replaced with the official Dataset doi.

## 9. Acknowledgments

The author is grateful to anonymous reviewers for their invaluable comments.

The research is supported by a grant of the Russian Science Foundation (project no. 22-71-00075)

## 10. Author Contributions

I.K. conceived and designed the research, analysed the results, and wrote and revised the manuscript.

## 11. Competing interests

The author declares no competing interests.

## 12. Appendix

### Appendix A. Mean-field approximation for the Advanced Model (system (2))

All the computations presented in this Section of the Appendix completely follow those from (Kozitsin, 2022) (see Appendix 3 in that paper).

First, let us write down the probabilities that the population of ordinary agents holding opinion $x_f$ will increase and decrease by one in the next time moment correspondingly (essentially, the assumption on the fully mixed population is employed here):

$$\Pr\left[Y_f^y(t+1) = Y_f^y(t) + 1\right] = \sum_{s,l,k} \frac{Y_s^y(t)}{N}\left(1 - \delta_{s,f}\right)\left[\frac{Y_l^y(t) - \delta_{s,l}}{N}p_{s,l,k}^y + \frac{Y_l^u(t)}{N}p_{s,l,k}^u\right]\delta_{k,f},$$

$$\Pr\left[Y_f^y(t+1) = Y_f^y(t) - 1\right] = \sum_{s,l,k} \frac{Y_s^y(t)}{N}\delta_{s,f}\left[\frac{Y_l^y(t) - \delta_{s,l}}{N}p_{s,l,k}^y + \frac{Y_l^u(t)}{N}p_{s,l,k}^u\right]\left(1 - \delta_{k,f}\right).$$

Next, suppose that agents' opinions at time $t$ are specified. Then, the mathematical expectation of the number of agents espousing opinion $x_f$ at time $t+1$ can be obtained as:

$$\mathrm{E}\big[Y_f^y(t+1)\big] = Y_f^y(t) + \Pr\big[Y_f^y(t+1) = Y_f^y(t)+1\big] - \Pr\big[Y_f^y(t+1) = Y_f^y(t)-1\big]$$

or

$$\mathrm{E}\big[Y_f^y(t+1)\big] = Y_f^y(t) + \sum_{s,l,k} \frac{Y_s^y(t)}{N}\left[\frac{Y_l^y(t)-\delta_{s,l}}{N}p_{s,l,k}^y + \frac{Y_l^u(t)}{N}p_{s,l,k}^u\right]\big(\delta_{k,f}-\delta_{s,f}\big).$$

Now it is time to recall that the number of agents (of both types) in the system is huge and, importantly, opinions of stubborn agents (which are exogenous to the model) change no faster than those of ordinary agents. As a result, random variable $Y_f^y(t+1)$ can be safely approximated as

$$Y_f^y(t+1) = Y_f^y(t) + \sum_{s,l,k} \frac{Y_s^y(t)}{N}\left[\frac{Y_l^y(t)}{N}p_{s,l,k}^y + \frac{Y_l^u(t)}{N}p_{s,l,k}^u\right]\big(\delta_{k,f}-\delta_{s,f}\big). \qquad (A1)$$

Expression (A1) can be rewritten in the scaled time ($\tau = \frac{t}{N}, \delta\tau = \frac{1}{N}$):

$$Y_f^y(\tau+\delta\tau) = Y_f^y(\tau) + \sum_{s,l,k} \frac{Y_s^y(\tau)}{N}\left[\frac{Y_l^y(\tau)}{N}p_{s,l,k}^y + \frac{Y_l^u(\tau)}{N}p_{s,l,k}^u\right]\big(\delta_{k,f}-\delta_{s,f}\big)$$

or

$$\frac{y_f(\tau+\delta\tau)-y_f(\tau)}{\delta\tau} = \sum_{s,l,k} y_s(\tau)\big[y_l(\tau)p_{s,l,k}^y + u_l(\tau)p_{s,l,k}^u\big]\big(\delta_{k,f}-\delta_{s,f}\big).$$

For large $N$, one can obtain:

$$\frac{dy_f(\tau)}{d\tau} = \sum_{s,l,k} y_s(\tau)\big[y_l(\tau)p_{s,l,k}^y + u_l(\tau)p_{s,l,k}^u\big]\big(\delta_{k,f}-\delta_{s,f}\big).$$

## Appendix B. Proof of Statement 1

The right part (3) of equation (2) is a polynomial. As such, one can guarantee that the Cauchy problem (2), (4) has a unique solution $\vec{y}(\tau)$ on some interval. Let us first prove that $\vec{y}(\tau) \geq \vec{0}$. This property follows directly from one of the Comparison theorem (Hartman, 2002). Let us consider the auxiliary function $\vec{\psi}(\tau) \equiv \vec{0}$. For this function, one can write:

$$\vec{0} = \vec{\psi}(t_0) \leq \vec{y}(t_0)$$

and

$$\vec{0} \equiv \vec{\psi}' \leq \vec{f}(\vec{\psi}) \equiv \vec{0}.$$

As such, one can guarantee that $\vec{\psi}(\tau)$ is no greater than the upper solution to the Cauchy problem (2), (4). In the considered case, the upper solution is exactly $\vec{y}(\tau)$. Hence:

$$\vec{y}(\tau) \geq \vec{\psi}(\tau) \equiv \vec{0}.$$

Because $\vec{y}(\tau)$ is nonnegative, it is straightforward to see that this function is bounded above (this property comes from the fact that the sum of $\vec{y}(\tau)$'s components is constant). Hence, this solution can be extended on the whole axis.

## Appendix C. Derivation of system (10)

Let us obtain the explicit formula for the components $a_{ij}$ of matrix $A$ from system (10). First, calculate the partial derivative of the Hamiltonian (see (7)) with respect to $y_i$:

$$\frac{\partial H}{\partial y_i} = \left( \sum_j \lambda_j f_j - g \right)_{y_i} = \left( \sum_j \lambda_j \left\{ \sum_{s,l} y_s [y_l p^y_{s,l,j} + u_l p^u_{s,l,j}] - y_j \right\} - \sum_j w_j y_j \right)_{y_i}$$

$$= \left( \sum_{j,s,l} \lambda_j y_s y_l p^y_{s,l,j} + \sum_{j,s,l} \lambda_j y_s u_l p^u_{s,l,j} - \sum_j (\lambda_j + w_j) y_j \right)_{y_i}$$

$$= \sum_{j,l} \lambda_j y_l p^y_{i,l,j} + \sum_{j,s} \lambda_j y_s p^y_{s,i,j} + \sum_{j,l} \lambda_j u_l p^u_{i,l,j} - (\lambda_i + w_i)$$

$$= \sum_j \lambda_j \left[ \sum_s y_s p^y_{s,i,j} + \sum_l (y_l + u_l) p^u_{i,l,j} \right] - (\lambda_i + w_i)$$

$$= \sum_j \lambda_j \left[ \sum_s y_s p^y_{s,i,j} + \sum_l (y_l + u_l) p^u_{i,l,j} - \delta_{i,j} \right] - w_i.$$

After that, one should notice that $a_{ij}$ is a multiplier near $\lambda_j$ in $-\frac{\partial H}{\partial y_i}$. This observation finishes the derivation:

$$a_{ij}(\vec{y}, \vec{u}) = \delta_{i,j} - \sum_s y_s p^y_{s,i,j} - \sum_l (y_l + u_l) p^u_{i,l,j}.$$

## Appendix D. Proof of Statement 3.

For convenience, instead of analysing system (10), it is suggested to consider the following "inverted system"

$$\frac{d\vec{z}}{d\tau} = B\vec{z} - \vec{w},$$
$$\vec{z}(T_0) = -\vec{v}.$$

(A2)

in the interval $[T_0, T_1]$ where $T_0 = -t_1, T_1 = -t_0$ and $B = -A$ (the initial value $-\vec{v}$ is the same as the one in (10)). A straightforward observation is that the solution to system (10) can be obtained from that of (A2) after inverting the time. As such, if one proves that the solution of (A2) (which exists and is unique because for given $\vec{u}$ and $\vec{y}$ the Cauchy problem (A2) satisfies the Picard–Lindelöf theorem) is nonpositive, then it automatically follows that $\vec{\lambda}(\tau) \leq \vec{0}$.

To prove that $\vec{z}(\tau)$ is nonpositive, let us apply again the comparison theorem. As in Appendix B, the function $\vec{\psi}(\tau) \equiv \vec{0}$ will be used. For now, one can estimate:

$$\vec{0} = \vec{\psi}(T_0) \geq \vec{z}(T_0) = -\vec{v},$$

$$\vec{0} \equiv \vec{\psi}' \geq B\vec{\psi} - \vec{w} \equiv -\vec{w}.$$

From these inequalities, one concludes that

$$\vec{z}(\tau) \leq \vec{\psi}(\tau) \equiv \vec{0}$$

or

$$\vec{\lambda}(\tau) \leq \vec{0}.$$

The nonpositiveness of partial derivatives $\frac{\partial H}{\partial u_i}$ immediately follows from inequalities $\vec{\lambda}(\tau) \leq \vec{0}, \vec{y}(\tau) \leq \vec{0}$ because

$$\frac{\partial H}{\partial u_i} = \sum_j \lambda_j \sum_s y_s p_{s,i,j}^u.$$

## Appendix E. Singular control

Let us first assume that $d = 2$ partial derivatives, say $\frac{\partial H(\tau)}{\partial u_i}$ and $\frac{\partial H(\tau)}{\partial u_j}$, have the highest value in interval $I_\tau$. In this case, one can write:

$$\frac{d}{d\tau}\frac{\partial H(\tau)}{\partial u_i} = \frac{d}{d\tau}\frac{\partial H(\tau)}{\partial u_j}$$

or

$$\sum_{r,q}\left(\frac{d\lambda_r}{d\tau}y_q^* + \lambda_r\frac{dy_q}{d\tau}\right)\left(p_{q,i,r}^u - p_{q,j,r}^u\right) = 0.$$

Derivatives $\frac{dy_q}{d\tau}$ and $\frac{d\lambda_r}{d\tau}$ can be expressed via (2) and (10):

$$\sum_{r,q}\left(y_q^*\left\{w_r - \sum_k \lambda_k\left[\sum_s y_s^* p_{s,r,k}^y + \sum_l (y_l^* + u_l^*)p_{r,l,k}^u - \delta_{r,k}\right]\right\}\right.$$

$$\left. + \lambda_r\left\{\sum_{s,l} y_s^*\left[y_l^* p_{s,l,q}^y + u_l^* p_{s,l,q}^u\right] - y_q^*\right\}\right)\left(p_{q,i,r}^u - p_{q,j,r}^u\right) = 0,$$

or

$$\sum_{r,q}\left(y_q^*\left\{w_r - \sum_k \lambda_k\left[\sum_s y_s^* p_{s,r,k}^y + \sum_l y_l^* p_{r,l,k}^u - \delta_{r,k}\right]\right\} + \lambda_r\left\{\sum_{s,l} y_s^* y_l^* p_{s,l,q}^y - y_q^*\right\}\right)\left(p_{q,i,r}^u - p_{q,j,r}^u\right) + \sum_{r,q,k,l} u_l^*\left[\lambda_r y_q^* p_{k,l,q}^u - y_q^* \lambda_k p_{r,l,k}^u\right]\left(p_{q,i,r}^u - p_{q,j,r}^u\right) = 0. \quad\text{(A3)}$$

In this particular case, let us set $u_l^* = 0$ for $l \notin \{i,j\}$ and then use (A3) to estimate $u_i^*$ and $u_j^*$:

$$C^{ij} + c_i^{ij}u_i^* + c_j^{ij}u_j^* = 0, \quad\text{(A4)}$$

where

$$C^{ij} = \sum_{r,q}\left(y_q^*\left\{w_r - \sum_k \lambda_k\left[\sum_s y_s^* p_{s,r,k}^y + \sum_l y_l^* p_{r,l,k}^u - \delta_{r,k}\right]\right\}\right.$$

$$\left. + \lambda_r\left\{\sum_{s,l} y_s^* y_l^* p_{s,l,q}^y - y_q^*\right\}\right)\left(p_{q,i,r}^u - p_{q,j,r}^u\right)$$

and

$$c_i^{ij} = \sum_{r,q,k}\left[\lambda_r y_q^* p_{k,i,q}^u - y_q^* \lambda_k p_{r,i,k}^u\right]\left(p_{q,i,r}^u - p_{q,j,r}^u\right), c_j^{ij}$$

$$= \sum_{r,q,k}\left[\lambda_r y_q^* p_{k,j,q}^u - y_q^* \lambda_k p_{r,j,k}^u\right]\left(p_{q,i,r}^u - p_{q,j,r}^u\right).$$

Note that in (A4) one of the variables $u_i^*$ and $u_j^*$ is redundant because $u_i^* + u_j^* = n^u$. As a result, one comes to the following equation:

$$\left(c_i^{ij} - c_j^{ij}\right)u_i^* = -C^{ij} - c_j^{ij}.$$

The situation when $d > 2$ is elaborated analogously: if there are $d$ partial derivatives $\frac{\partial H(\tau)}{\partial u_{i_1}}, \dots, \frac{\partial H(\tau)}{\partial u_{i_d}}$ with the highest value in interval $I_\tau$, then to maximize the Hamiltonian, one should set $u_i^* = 0$ for $i \notin \{i_1, \dots, i_d\}$ and find the values of $u_{i_1}^*, \dots, u_{i_d}^*$ from the system

$$\begin{bmatrix} c_{i_1}^{i_1 i_2} - c_{i_d}^{i_1 i_2} & \cdots & c_{i_{d-1}}^{i_1 i_2} - c_{i_d}^{i_1 i_2} \\ \vdots & \ddots & \vdots \\ c_{i_1}^{i_{d-1} i_d} - c_{i_d}^{i_{d-1} i_d} & \cdots & c_{i_{d-1}}^{i_{d-1} i_d} - c_{i_d}^{i_{d-1} i_d} \end{bmatrix} \begin{bmatrix} u_{i_1}^* \\ \vdots \\ u_{i_{d-1}}^* \end{bmatrix} = \begin{bmatrix} -C^{i_1 i_2} - c_{i_d}^{i_1 i_2} \\ \vdots \\ -C^{i_{d-1} i_d} - c_{i_d}^{i_{d-1} i_d} \end{bmatrix}$$

and using the equality

$$u_{i_1}^* + \cdots + u_{i_{d-1}}^* + u_{i_d}^* = n^u.$$

## Appendix F. Technical details of the algorithm

Here, some technical details regarding the implementation of the FBS method are presented. It was realized in JupiterHub using the Python 3 language, as well as the numerical experiments. This Subsection of the Appendix focuses on **Steps 2–4** of the FBS method that needs a more comprehensive description.

### Step 2

In this step, the algorithm solves the Cauchy problem (2), (4). Here, the Python function "solve_ivp" from the module "scipy.integrate" is used[2]. This function aims at solving the Cauchy problem numerically and affords an opportunity to use different integration methods. I started with the explicit Runge-Kutta method of order 5(4). However, preliminary experiments revealed that the quality of approximation can be improved by implementing its implicit version from the Radau IIA family (Hairer *et al*., 1991). It was also shown that for some transition matrices and in long integration intervals, the Runge-Kutta method had built solutions that violated the restriction $y_1 + \cdots + y_m = n^y$. Typically, it is reflected in the fact that all the state variables tend to zero but, in some cases, they may grow unexpectedly. To exclude these situations, the Runge-Kutta method was slightly modified (hereafter – MRK (modified Runge-Kutta) method, see Figure A1). The MRK method elaborates the solution using the sequence of standard Runge-Kutta runs. Each run goes until the restriction holds. At the time of the first violation of the restriction, the Runge-Kutta run stops and an ad-hoc correction is applied on this "inaccurate" vector (if necessary, the incorrect vector is pushed towards the area of nonnegative values, and then, if the sum of its component differs from $n^y$ by more than $10^{-5}$, is normalized to ensure that the sum is equal to $n^y$). As a result, a new, corrected vector is obtained that serves as an initial condition for the next Runge-Kutta run, and so forth, until all the grid points are not filled. Due to the organization of the MRK method, in each time step, it obtains at least one new estimated value on the grid, and thus the MRK method always converges. However, except for some extreme cases (see Scenario 2 in Subsection 5.1), the MRK method typically needs only a few standard Runge-Kutta runs to obtain the solution, instead of casting the correction procedure on each grid point. In the case of Scenario 2, if one considers the transition matrix (S2), then problem (2), (4) becomes stiff, and numerical methods are hardly applicable here.

---

[2] https://docs.scipy.org/doc/scipy/reference/generated/scipy.integrate.solve_ivp.html#r179348322575-1

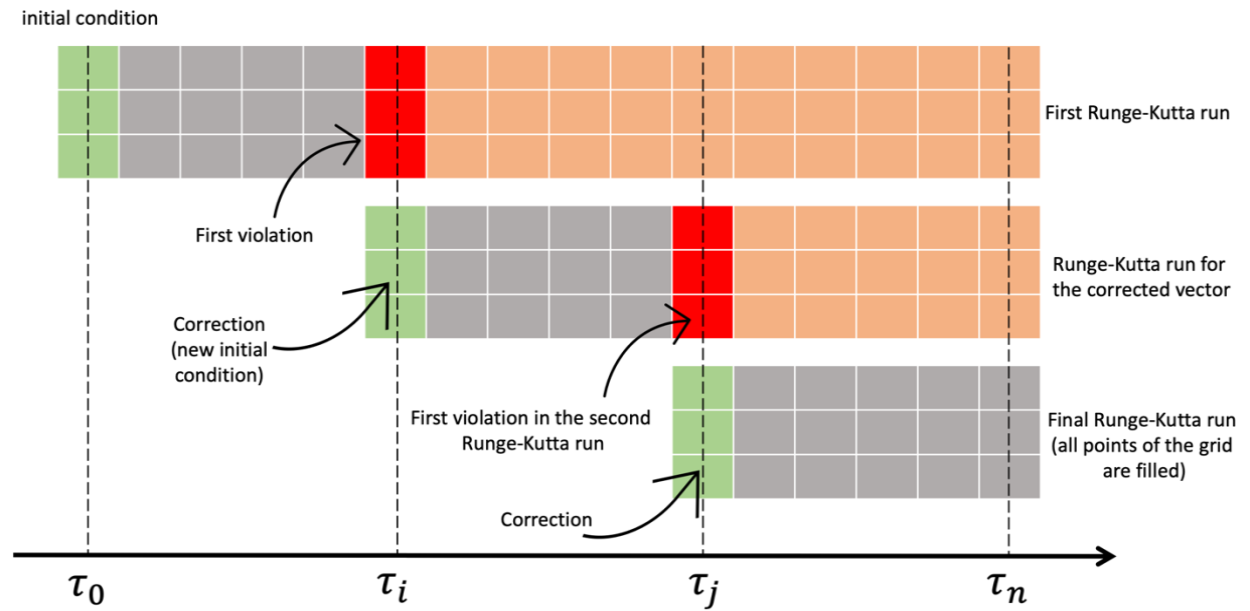

Figure A1. Illustration of how the MRK method operates. In this example, the algorithm needs three standard Runge-Kutta runs to approximate the solution to the Cauchy problem. Within the first run, the MRK method obtains the solution in the whole interval $[\tau_0, \tau_n]$, but only the values from the span $[\tau_0, \tau_{i-1}]$ are valid (grey colour). Next, the MRK method takes the first invalid vector (obtained on the point $\tau_i$, depicted with red colour), corrects it, and uses it in the second Runge-Kutta run as a new initial point (green colour). By doing so, the MRK method obtains a subsequent piece of the general solution in the interval $[\tau_i, \tau_{j-1}]$. The third, final, Runge-Kutta run achieves the solution for $\tau \in [\tau_j, \tau_n]$. Finally, the general solution is obtained via the merging of these three subsequent approximations.

*Step 3*

In Step 3, the algorithm solves the Cauchy problem (10). Because the Python function "solve_ivp" assumes that the initial condition is posed on the left endpoint of the interval, this function can be applied directly to the Cauchy (10). In this regard, instead of system (10), the algorithm considers the "inverted system" (A2). Note that there is no need to use the MRK method here because there are no restrictions on the adjoint functions (except for the case when one knows that they are nonpositive (see Statement 3) – this theoretically-derived property was approved in numerical experiments).

### Step 4

In this Step, problems can arise only if elaborating the case of singular control. First, if the determinant of the matrix

$$C = \begin{bmatrix} c_{i_1}^{i_1 i_2} - c_{i_d}^{i_1 i_2} & \cdots & c_{i_{d-1}}^{i_1 i_2} - c_{i_d}^{i_1 i_2} \\ \vdots & \ddots & \vdots \\ c_{i_1}^{i_{d-1} i_d} - c_{i_d}^{i_{d-1} i_d} & \cdots & c_{i_{d-1}}^{i_{d-1} i_d} - c_{i_d}^{i_{d-1} i_d} \end{bmatrix}$$

is equal to zero, then there could be no solutions or, alternatively, infinitely many. Further, even if the matrix is not singular, solutions could be potentially negative, or their sum may exceed $n^b$. These situations are as follows. If matrix $C$ is singular, then a minimization procedure is run in which one can try to find the vector $[u_{i_1} \quad \cdots \quad u_{i_{d-1}}]^T$ that minimizes the squared difference between the left and right parts of (12) and, at the same time, meet restrictions (A5):

$$\begin{aligned} 0 &\leq u_{i_k}, \ k \in [d]; \\ u_{i_1} + \cdots + u_{i_{d-1}} &\leq n^u. \end{aligned} \tag{A5}$$

If $C$ is nonsingular but the solution of (12) violates restrictions (A5), then one index $i_k$ from set $\{i_1, \dots, i_d\}$ is chosen at random and the corresponding component of the control vector is defined as $u_{i_k}(\tau_k) = n^u$. However, at least in numerical experiments presented in this paper, matrix $C$ was always nonsingular and the corresponding solutions were valid.